\documentclass[useAMS,usenatbib]{mn2e}
\usepackage{graphicx}
\usepackage{amsmath}

\def\gtsima{$\; \buildrel > \over \sim \;$}
\def\ltsima{$\; \buildrel < \over \sim \;$}
\def\gtrsim{\lower.5ex\hbox{\gtsima}}
\def\lesssim{\lower.5ex\hbox{\ltsima}}

\title[Non-parametric classification of a BAT sample]{The merger fraction of active and inactive galaxies in the local Universe through an improved non-parametric classification}
\author[Cotini et al.]{Stefano Cotini,$^{1,2}$\thanks{E-mail: stefano.cotini@gmail.com}
Emanuele Ripamonti,$^{2}$\thanks{E-mail: ripamonti.e@gmail.com}
Alessandro Caccianiga,$^{1}$
Monica Colpi,$^{2}$
\newauthor
Roberto Della Ceca,$^{1}$
Michela Mapelli,$^{3}$
Paola Severgnini$^{1}$
and Alberto Segreto$^{4}$
\\
$^{1}$INAF-Osservatorio Astronomico di Brera, Via Brera 28, I-20121, Milano, Italy \\
$^{2}$Universit\`{a} degli Studi di Milano-Bicocca, Dipartimento di Fisica \textquotedblleft G. Occhialini\textquotedblright, Piazza della Scienza 3, I-20126, Milano, Italy \\
$^{3}$INAF-Osservatorio Astronomico di Padova, Vicolo dell'Osservatorio 5, I-35122, Padova, Italy \\
$^{4}$INAF-Istituto di Astrofisica Spaziale e Fisica Cosmica, Via U. La Malfa 153, I-90146, Palermo, Italy
}
\begin{document}

\date{}

\pagerange{\pageref{firstpage}--\pageref{lastpage}} \pubyear{2012}

\maketitle

\label{firstpage}

\begin{abstract}
We investigate the possible link between mergers and the enhanced
activity of supermassive black holes (SMBHs) at the centre of
galaxies, by comparing the merger fraction of a local sample ($0.003
\leq z < 0.03$) of active galaxies - 59 active galactic nuclei (AGN)
host galaxies selected from the all-sky \textit{Swift} BAT
(\textit{Burst Alert Telescope}) survey - with an appropriate control
sample (247 sources extracted from the Hyperleda catalogue) that has
the same redshift distribution as the BAT sample.  We detect the
interacting systems in the two samples on the basis of non-parametric
structural indexes of concentration ($C$), asymmetry ($A$), clumpiness
($S$), Gini coefficient ($G$) and second order momentum of light
($M_{20}$). In particular, we propose a new morphological criterion,
based on a combination of all these indexes, that improves the
identification of interacting systems. We also present a new software
- \textsc{PyCASSo} (Python $CAS$ Software) - for the automatic
computation of the structural indexes.  After correcting for the
completeness and reliability of the method, we find that the
  fraction of interacting galaxies among the active population
  ($20_{-5}^{+7}$ per cent) exceeds the merger fraction of the control
  sample ($4_{-1.2}^{+1.7}$ per cent). Choosing a mass-matched control
  sample leads to equivalent results, although with slightly lower
  statistical significance. Our findings support the scenario in which
  mergers trigger the nuclear activity of supermassive black holes.
\end{abstract}


\section{Introduction} \label{sec:intro}
Observations indicate that the growth history of supermassive black holes (SMBHs, $M_{\rm BH} > 10^6$ M$_{\odot}$) is closely connected to that of their host galaxies.  The discovery of scaling relations, linking the black hole mass to properties of the host in the local Universe, hints for a scenario of galaxy-SMBH symbiotic evolution (\citealt{magorrian98}; \citealt{ferraresemerritt00}; \citealt{gebhardt00}; \citealt{marconihunt03}; \citealt{haringrix04}; \citealt{ferrareseford05}; \citealt{graham12a}; \citealt{graham12b}). 
In particular, the near ubiquity of SMBHs in massive spheroids indicates that black hole growth, mainly driven by gas accretion (e.g. \citealt{marconi04}; \citealt{croton06}; 
\citealt{volonteri12}; \citealt{merloni12}), is favored in galaxies where the importance of organized rotation both in the gaseous and stellar component is weak.
As morphological properties of galaxies are likely to be determined by their complex assembly history and can be transient features, the processes that determine the formation and evolution of galaxies affect hand in hand the formation and evolution of  SMBHs, and in particular their fueling.

Theoretical models indicate that galaxy formation and evolution is driven by accretion of gas from the cosmic environment (e.g. \citealt{keres05}; \citealt{bournaud2005}; \citealt*{mapelli08}; see \citealt{sancisi08} for a review) and by halo-halo interactions 
both involving multiple minor mergers or major galaxy-galaxy mergers (e.g. \citealt{white78}; \citealt{millersmith80}; \citealt{gerhard81}; \citealt{negrowhite83}; \citealt{lakedressler86}; \citealt{barnes88}; \citealt{dekel09}; \citealt{bournaud11}; see \citealt{mirabel01} for a review). 
More recently, the mode of gas accretion has been recognized as playing a potentially critical role in shaping galaxies (\citealt{sales12}), leaving open the possibility that spheroids form via multiple episodes of misaligned gas inflows, besides major mergers.
In lack of a broad consensus, observations of AGNs and of their galaxy hosts, from suitably selected samples, can provide clues on the mechanisms triggering the SMBH activity, and on their coevolution.  

A longstanding issue is how the gas can lose enough angular momentum from the large scale ($\sim 0.1-100$ kpc) down to the SMBH's horizon scale ($\sim 10^{-5}$ pc). 
A possible scenario involves gravitational perturbations due to tidal interactions between galaxies in close fly-bys (on $\sim 10-70$ kpc scales) or/and violent galaxy mergers occurring on smaller scales of  $\sim$ kpc or less.
These perturbations may drive large quantities of gas towards the centre of the merger remnant  (e.g., \citealt{kauffmannhaehnelt00}; \citealt{springel05}; \citealt{hopkins2006}).  This accumulated gas may induce both an intense starburst phase and an enhanced nuclear activity (active SMBH), whose feedback, in turn, can act as a mechanism to regulate subsequent star-formation and accretion (\citealt{churazov01}; \citealt{best06}; \citealt{schawinski06}; \citealt{schawinski07}; \citealt{mcnamaranulsen07}). Galaxy interactions/mergers should be therefore responsible not only for large scale ($\gtrsim{}10^3$ pc) 
morphological distortions, but also for the inflow of gas down to the typical scale of SMBH accretion ($\lesssim{}10^{-4}$ pc).

If SMBH activity is triggered, at least partially, by galaxy mergers, the fraction of galaxies with clear sign of being the results of 
interactions/mergers should be statistically higher in a sample of AGN--host galaxies than in a sample of field galaxies.
This and other similar observational tests have been carried out in the last few years with somehow contrasting results (see e.g. \citealt{petrosian82}; \citealt{dahari84}; \citealt{dahari85}; \citealt{keel85}; \citealt{fuentesstocke88}; \citealt*{virani00}; \citealt{schmitt01}; \citealt{miller03}; \citealt{grogin05}; \citealt{waskett05}; \citealt{koulouridis06}; \citealt{serber06}). 
In particular, while some studies claim a connection between nuclear activity and the presence of close
companions or tidal distortions (e.g., \citealt{dahari84}; \citealt{keel85}; \citealt*{rafanelli95}; \citealt{koss2010}; \citealt{koss11}; \citealt{koss2012}; \citealt{ramosalmeida11}; \citealt{silverman11}; \citealt{ellison11}; \citealt*{liu12}), other studies indicate that there is statistically little support to a AGN-merger connection (\citealt*{barton00}; \citealt{schmitt01}; \citealt{dunlop03}; \citealt{grogin05}; \citealt{coldwelllambas06}; \citealt{alonso07}; \citealt{li08}; \citealt{ellison08}; \citealt{gabor2009}; \citealt{darg10}; \citealt{cisternas2011}; \citealt{kocevski12}).

The differences between various studies might be due to biases in the choice of the galaxy sample. For example, obscured AGNs can be missed in studies based on optical emission-line ratios, optical spectral classification or even soft X-ray fluxes. Among the aforementioned studies, only \citet{koss2010} use a sample of hard X-ray selected AGNs, and find a strong excess of merging systems with respect to a control sample.

Another source of error is counting chance superposition galaxy pairs as physically interacting galaxies 
(for more details about this source of error we refer the reader to section 6.1 of \citealt{ellison11}).  

The third source of bias is the possible time delay between the merger and the switch on of the nuclear activity. Various studies (e.g. \citealt{ellison08}; \citealt{schawinski09}; \citealt{schawinski10}, and references therein) find empirical evidences that  mergers enhance star formation first, and only at later epochs trigger the AGN phase ($\sim{}500$ Myr after the starburst). 
In fact \citet*{smirnova10} analyse a sample of apparently isolated Seyfert galaxies and find that about 35 per cent of them show tidal tails, consistent with a gas-rich merger (likely a minor merger) in the last $0.5-1$ Gyr. 
Thus, samples of galaxy pairs might miss, by default, late merger phases and gas-rich minor mergers.
This problem is less acute when empirical measures of galaxy
morphology are used, as they can identify a galaxy as the result of an
interaction/merger even when it lacks a companion (provided that
interaction features are strong enough). Therefore, these measures are
sensitive both to the initial and the late stages of mergers, and are
less biased against specific merger phases.

In this paper we re-address the possible link between mergers/interactions (in the following, we will use the two terms as synonimous) and SMBH activity, by comparing the merger fraction of an AGN host galaxy sample to the typical merger fraction of galaxies in the local Universe. 

To satisfy the need that both the galaxy sample and the method of analysis are as unbiased as possible, (i) we use a hard ($>10$ keV) X-ray selected AGN sample (not to miss obscured sources, with the partial exception of the heavily absorbed Compton thick AGNs, i.e. those sources with absorbing column densities exceeding $10^{24}$ cm$^{-2}$), and (ii) we adopt a non-parametric morphological analysis (to identify truly interacting galaxies even in late merger phases).

Moreover, we propose an improved technique for evaluating the merger fraction of a galaxy sample by using a method that is objective, reliable and fast, so that it can be applied, in the future, to larger samples of galaxies; we also define the completeness and the reliability coefficients, that allow a statistical correction of the merger fraction and further reduce possible residual errors in the automated classification.

This paper is organized as follows: Section \ref{sec:data} presents the galaxy samples and the procedure adopted for their unbiased selection; Section \ref{sec:data-analysis} explains the non-parametric morphological method used for the analysis; Section \ref{sec:results} presents our estimates of the merger fraction of the AGN BAT sample and of the control sample; Section \ref{sec:conclusions} outlines a summary of the most important points. Appendices \ref{app:data-processing} and \ref{app:image-degradation} present respectively the data processing algorithms (including a detailed description of the software that we developed for our automated classification) and a discussion on  the image degradation effects that affect data analysis.

\section{Sample selection} \label{sec:data}

The aim of this work is to study the possible link between mergers and SMBH activity, by comparing the merger fraction of an AGN host galaxy sample to the typical merger fraction of galaxies in the local Universe. To this purpose, we select two samples: 
the first one is a hard (15-195 keV) X-ray selected sample of active galaxies (will be addressed here as the BAT sample), which is similar - with several objects in common - to the sample already used in  \citet{koss2010}. 
The second one is an optically selected control sample of galaxies (without any imposition on their active nature) that we extract from the Hyperleda catalogue (\citealt{paturel2003}).
We impose on both samples a minimum redshift of 0.003, to avoid too extended sources (image processing faces some difficulties in these cases) and a maximum redshift of 0.03, because the optical counterparts of the selected galaxies need to match the requirements for our morphological analysis (see Appendix \ref{app:pycasso}).

\subsection{BAT sample} \label{sec:bat}
The \textit{Burst Alert Telescope} (BAT) is a coded aperture imaging camera on-board the \textit{Swift} satellite \citep{gehrels2004}; it has a wide field of view (1.4 steradian), a PSF of 17 arcmin (FWHM) and it operates in the 15-195 keV energy range. 
To select a sample of AGNs out of \textit{Swift} BAT observations, we adopt the \textit{Palermo Swift-BAT Hard X-ray catalogue} \citep{palermo}, that collects the data relative to the first 54 months of the \textit{Swift} mission and is therefore one of the most complete, well defined and extended catalogues of hard X-ray sources up to date. It contains 1256 sources with a signal to noise ratio greater than 4.8, a flux limit of $6.0\times 10^{-12}$ erg cm$^{-2}$ s$^{-1}$ and a counterpart identification with a 95 per cent confidence level. This catalogue represents a relatively unbiased sample of AGNs, because it is based on a particular hard X-ray band, where biases against absorbed AGNs are less important.
For our analysis, we extract from this catalogue a complete sample of 523 sources, with absolute Galactic latitude $|b| > 15^{\circ}$, $S/N > 5$ and flux greater than $8.0 \times 10^{-12}$ erg cm$^{-2}$ s$^{-1}$. 
Second, we select a complete sub-sample in the redshift interval $0.003 \leq z < 0.03$ and, finally, we restrict to the area of sky covered by the Sloan Digital Sky Survey Data Release 8 (http://www.sdss3.org/dr8/), to make use of the optical data offered by this survey. The final BAT active galaxy sample\footnote{This sample does not include two sources that have too low resolution for being analysed and one source that is very close to a bright star, which invalidates our analysis.} consists of 59 sources (15 at redshift $0.003 \leq z<0.01$, 16 at redshift $0.01\leq z<0.02$ and 28 at redshift $0.02\leq z<0.03$), which represent $\sim35$ per cent of the total number of galaxies belonging to the complete sample in the same redshift interval $0.003-0.03$ (169 objects).\\
The BAT sample is not a mere selection of galaxies, but of systems instead: the sources are selected on the basis of the presence of one AGN at least, but the poor angular resolution of \textit{Swift} BAT observations does not allow to distinguish the possible X-ray emission of multiple AGNs in pairs or group of galaxies. 
As a consequence, in case of merging galaxies, the ensemble of objects is considered as a single (interacting) system, likewise each isolated galaxy represents a single (but non interacting) system. In particular, the \textquotedblleft interacting\textquotedblright{ }or \textquotedblleft non interacting\textquotedblright{ }classification is determined from the results of the automated structural analysis (see Section \ref{sec:data-processing}).

\subsection{The control sample} \label{sec:control-sample}
The control sample is used to evaluate the average merger fraction among galaxies and to compare it with the same value found in the BAT sample (i.e. among AGNs), so it has to match the redshift distribution of the BAT sample and it must be unbiased towards interacting or isolated systems. \\
For example, a random sampling among SDSS galaxies would lead to an overestimate of the merger fraction, because interacting systems have more chances to be selected than isolated galaxies (in fact they can be sorted out by each one of their members).
Therefore, we replicate the particular \textquotedblleft system classification\textquotedblright{ }of the BAT sample also in the control one.
In the following we describe the procedure used to define the control sample:
\begin{itemize}
\item[-] We select three random square boxes of sky fully covered by SDSS imaging. All boxes have a side of 7.5 degrees and contain, on average, $\sim300$ galaxies\footnote{We point out that, due to our subselections and the impossibility to analyse all the images, the number of valid objects in each box is usually reduced almost by 20 per cent.} in the \mbox{$0.003 \leq z<0.03$} redshift interval. The choice of multiple medium-size boxes, instead of a single large box, avoids biases related to local peculiar environment (i.e. galaxy groups or clusters). The size of the boxes ensures a significant number of sources inside each one, so that possible border effects become unimportant (i.e. the loss of one galaxy of a pair that lies halfway the edge of the box).
\item[-] We consider all the sources in the Hyperleda catalogue present in the three boxes of sky quoted above. For each galaxy, we acquire the SDSS image and, on the basis of the structural parameters (asymmetry, clumpiness, Gini coefficient, second order momentum of light - see Section \ref{sec:data-analysis}), we distinguish whether it is interacting or isolated.
\item[-] We switch from the \textquotedblleft galaxy classification\textquotedblright{ } to the \textquotedblleft system classification\textquotedblright: we consider as a multiple system every ensemble\footnote{The extent of the ensemble depends on the number and the kind of sources falling into the aperture (automatically computed on the basis of the light profile of the central galaxy, see Appendix \ref{app:pycasso}) for the estimation of structural indexes. However, in general, it is unlikely that galaxies with separation greater than 30 kpc are included in the same aperture.} of sources in which at least one galaxy has been identified as interacting through our classification. We consider only one galaxy of each system, so that each system is represented only by one entry.
At this point, the control sample consists of 734 systems (79 at redshift \mbox{$0.003 \leq z<0.01$,} 67 at redshift $0.01\leq z<0.02$ and 588 at redshift $0.02\leq z <0.03$). 
\item[-] The redshift distribution of these sources is considerably different from the BAT sample, so possible redshift-related effects (i.e. an evolution of the merger fraction) may alter significantly our comparison. For this reason, we reduce our sample by randomly extracting, in each redshift bin, the right number of sources to match the BAT sample's redshift distribution.
\end{itemize}

At the end of this procedure we obtain a redshift-matched control sample of 247 sources, distributed as shown in Table \ref{table:result-z-bin}, that are fully comparable with the BAT ones. We point out that the control sample contains both active and quiescent galaxies at random, because we want to check whether the merger fraction of the BAT AGN sample is significantly higher than the typical merger fraction in the local Universe.

\section{Data analysis} \label{sec:data-analysis}
In this work we aim to determine the merger fraction of two samples of galaxies using a method that is objective, reliable and fast, so that it can be applied, in the future, to larger samples of galaxies. 
To this end, different techniques have been developed in the last decades (i.e. \citealp{byun1995}; \citealp{abraham1996}; \citealp{lefevre2000}; \citealp{patton2000,patton2002}; \citealp{peng2002}; \citealp{conselice2003}; \citealp{blanton2003}; \citealp{lotz2004}; \citealp{scarlata2007}) but there is not yet a method that has been proven to be clearly superior to the others. \\
Pair counts require a strong observational effort, because they need redshift measurements for each galaxy, to avoid chance superpositions. Moreover, even pairs of galaxies at the same redshift could be not gravitationally bound, leading to an overestimate of the merger fraction. \\ 
Other techniques rely on the identification of galaxies that, due to gravitational interactions with a close companion, show morphological perturbations. The visual, qualitative, classification is the most used and accurate method, but it is intrinsically subjective and becomes less and less reliable with increasing redshift, because of the lower resolution and $S/N$ ratio. Moreover, it is time consuming, and, therefore, it is not applicable on very large samples of galaxies. \\
Quantitative classifications are less accurate but more objective, and allow corrections for high redshifts, because the image degradation is measurable. Among these, we can distinguish between parametric and non parametric classifications. In the first kind, the projected light distribution of the galaxy is either fitted as a whole with an analytical model (like the S\'ersic or the de Vaucouleurs profile), or it is split in its various components (i.e. a bulge and a disk), that are fitted separately. Nevertheless, these methods are quite unsuitable for irregular or disturbed galaxies and, in case of close pairs, the subtraction of the extra light coming from the companion is not trivial. Non parametric techniques are not based on any analytical models, so they are equally applicable on every kind of galaxy; however, it is more difficult to convert their values into physically meaningful results. An interesting, non parametric classification has been developed in recent years by \citet{conselice2003} and \citet{lotz2004}: it consists in a set of five structural indexes that measure specific properties of a galaxy. The first three parameters, concentration ($C$), asymmetry ($A$) and clumpiness ($S$), presented by \citet{conselice2003}, are referred as the $CAS$ system; the other two indexes, introduced by \citet{lotz2004}, are the Gini coefficient ($G$) and the second order momentum of light ($M_{20}$). We decide to adopt this non-parametric approach for our analysis and we will refer to the whole set of indexes as $CASGM$ system. 
As in a visual analysis, the $CASGM$ method becomes less reliable in case of low resolution or $S/N$ ratio, but these effects have been well quantified by \citet{lotz2004} and are reported at the end of in Appendix \ref{app:pycasso}. Taking into accounts this limits, we have imposed a maximum redshift of 0.03 to our samples, so that SDSS images ensure the minimum requirements for the automated analysis.

\subsection{The $CASGM$ parameters} \label{sec:casgm-parameters}
In order to compute these parameters, we need first to determine the extension of the galaxy, which is based on the Petrosian radius. The Petrosian index of a galaxy is the ratio between the mean surface brightness inside radius $R$, $\bar{\mu}(r<R)$, and the surface brightness $\mu(R)$ at $R$, that is: 
\begin{equation}
	\eta(R) = \frac{\bar{\mu}(r<R)}{\mu(R)} \, .
\label{eq:petrosian_index}
\end{equation}
The Petrosian radius is the radius $r_{\rm P}$ at which the inverted Petrosian index is equal to 0.2 \citep{petrosian1976}. 
For the $CAS$ system, the area of the galaxy is the circular area inside $1.5$ times the Petrosian radius at $r(\eta=0.2)$, with centre in the point that minimizes the asymmetry of the galaxy. 

\begin{itemize}
\item \textbf{Concentration:} the concentration index is the ratio of the light inside an inner aperture (circular or elliptical) to the light inside an outer aperture. The $CAS$ system adopts the \citet{bershady2000} definition, so $C$ is defined as:
	\begin{equation}
			C = 5 \,  \, log \left ( \frac{r_{80}}{r_{20}} \right ) \, ,
	\label{eq:C}
	\end{equation}
where $r_{80}$ and $r_{20}$ are the radii that contain the 80 per cent and the 20 per cent of the total light of the galaxy, respectively. Typical values of $C$ range from $\sim2$ to $\sim5$: elliptical galaxies and spheroidal systems usually have $C>4$, disk galaxies 		have $3<C<4$, while galaxies with a low surface brightness or a low velocity dispersion have $C\sim2$. \\

\item \textbf{Asymmetry:} the $A$ coefficient measures the asymmetry degree of the galaxy light distribution under a $180^{\circ}$ 	rotation. This index was originally used to describe galaxy morphologies (i.e \citealp{abraham1996}), but we follow the slightly modified formulation of \citet{conselice2000b}. This index is computed by subtracting the $180^{\circ}$ rotated image to the original one, and by normalizing the residuals by the total flux of the galaxy. This value is then corrected by subtracting the asymmetry contribution of the background (i.e. produced by a luminosity gradient or a close stellar halo), which is computed in the same way. Therefore, the final value of $A$ is
\begin{equation}
	\begin{split}
	A & = \frac{\sum_{i,j}|I(i,j)-I_{180}(i,j)|}{\sum_{i,j}|I(i,j)|}+ \\ 
		& -\frac{\sum_{i,j}|B(i,j)-B_{180}(i,j)|}{\sum_{i,j}|I(i,j)|} \, ,
	\end{split}
\label{eq:A}
\end{equation}
where $I$ and $B$ are respectively the original image of the galaxy and of the background, while $I_{180}$ and $B_{180}$ are their rotated images. This coefficient is sensitive to all the processes that introduce a certain degree of asymmetry in the light distribution, such as star forming regions, dust bands and mergers. The relative contribution of these elements have been studied by \citet{conselice2003}, who showed that small scale structures can make up only to the 30 per cent of the asymmetry of the galaxy; therefore $A$ is dominated by large scale effects and is a good tracer of mergers and gravitational distortions. \\

\item \textbf{Clumpiness:} the $S$ index has been introduced by \citet{conselice2003} to quantify the patchiness of the galaxy, that is the fraction of light coming from small scale structures, such as clumps of star formation. It is defined as the ratio of the flux contained in high frequency features to the total flux of the galaxy. It is computed by subtracting a blurred\footnote{The blurring is obtained by convolving the original image with a filter of width $\sigma = 0.25 r_{\rm P}$.} copy of the image to the original one, and then normalizing by the total flux of the galaxy. The value is then corrected by removing the background clumpiness, so it is equal to
	\begin{equation}
	\begin{split}
		S & = 10\frac{\sum_{i,j}(I(i,j)-I_{S}(i,j))}{\sum_{i,j}I(i,j)} + \\
		  & - \frac{\sum_{i,j}(B(i,j)-B_{S}(i,j))}{\sum_{i,j}I(i,j)} \, ,
	\end{split}
	\label{eq:S}
	\end{equation}
where $I$ and $B$ are the original image of the galaxy and of the background, respectively, while $I_{S}$ and $B_{S}$ are their blurred images. The nuclear $0.25 r_{\rm P}$ region is excluded from the computation, because it would give a high clumpiness contribution, which is not related to a region of young and intense star formation. Moreover, negative values after the subtraction of the smoothed image are forced to zero \citep{conselice2003}. \\
Large values of $S$ indicate that most of the light of the galaxy is accumulated in few and clumpy structures (i.e. starburst galaxies), while low values of $S$ indicate that the light distribution is smooth (i.e. elliptical galaxies).
\end{itemize}

$G$ and $M_{20}$ are based on the segmentation map of the galaxy defined by \citet{lotz2004}. In contrast with the circular and the elliptical apertures of the $CAS$ indexes, the segmentation map can assume any irregular shape, because its constraints (see Appendix \ref{app:pycasso-gini-momentum}) are only a brightness limit (to exclude the background and possible spurious pixels) and a continuity requirement (any source that is not directly connected with the galaxy is not taken into account). Therefore, the segmentation map can follow accurately the outline of the galaxy, especially in case of close couples and mergers.

\begin{itemize}
\item \textbf{Gini coefficient:} the Gini coefficient is a measure of statistical dispersion. It is usually adopted in economics to describe the inequality of a distribution (i.e. levels of income) and was adapted by \citet{abraham2003} and \citet{lotz2004} for the morphological classification of galaxies. The formulation of the Gini coefficient is based on the Lorentz curve 
\begin{equation}
	L(p) = \frac{1}{\bar{X}} \int_{0}^{p}{F^{-1}(u) \mathrm{d} u} \, ,
\end{equation}
where $p$ is the percentage of the faintest pixels, $F(x)$ the cumulative distribution function and $\bar{X}$ the average value of all the $X_i$ intensities. After some rearrangements \citep{glasser1962} and a correction to compensate for the Poissonian noise in the faintest regions of the galaxy \citep{lotz2004}, it can be expressed as 
\begin{equation}
	G = \frac{1}{\bar{|X|}n(n-1)}\sum^n_i(2i-n-1)|X_i| \, .
\label{eq:G}
\end{equation}
The Gini coefficient is computed on the segmentation map and represents a sort of generalized concentration index, in fact it tells whether the light is evenly distributed inside the galaxy, but does not depend on any particular centre. This index can range from zero, in case of a perfectly uniform distribution, to one, in case that all the light of the object is concentrated in a single pixel. \\

\item \textbf{Momentum of light:} the $M_{20}$ coefficient measures how far from centre are located the brightest pixels of the galaxy. It is based on the total second order momentum of light $M_{\rm tot}$, that is the sum, over all the pixels of the segmentation map, of the pixels' flux $f_i$ multiplied for its square distance from the centre:
\begin{equation}
	M_{\rm tot} = \sum_i^n M_i = \sum_i^n f_i \left [ (x_i - x_c)^2 + (y_i - y_c)^2 \right ] \, .
\end{equation}
The $x_c$ and $y_c$ variables are the coordinate of the galaxy centre, which is now defined as the pixel that minimizes the value of $M_{\rm tot}$. The $M_{20}$ coefficient is the second order momentum of the brightest 20 per cent of the galaxy.  To compute it, we follow the procedure in \citet{lotz2004}: the pixels of the segmentation map are sorted by decreasing flux; then the corresponding momenta $M_i$ are summed, until the sum of the brightest pixels equals 20 per cent of the total galaxy flux; finally this value is normalized by $M_{\rm tot}$, so
\begin{equation}
	M_{20} = {\rm log_{10}}\left(\frac{\sum_i M_i}{M_{\rm tot}}\right)  \quad {\rm while} \quad \sum_i f_i <  0.2 f_{\rm tot} \, .
\label{eq:M20}
\end{equation}
The normalization removes dependencies on the size of the object and its total flux, making $M_{20}$ less subject to inclination effects. Being weighted on the square of the distance from the centre, this index is especially suitable for detecting double nuclei systems (such as close galaxies in a merging phase), because the brightest pixels of the system are off-centre and they give a large contribution to the value of $M_{20}$.
\end{itemize}

The $CASGM$ system relies on the Petrosian radius, that, being based on a curve of growth, is independent of the galaxy size and largely insensitive to both the $S/N$ ratio and the surface brightness of the sources (see \citet{lotz2004} for a discussion about the influence of low $S/N$ and resolution on these parameters). \\
These indexes are related to galaxy morphologies, and the authors of the $CASGM$ system have calibrated a complete classification using the \citet{frei} catalogue. Moreover, linking couples of $CASGM$ indexes, they defined some fiducial sequences, that allow the separation of normal\footnote{Throughout the rest of this text, we will address to non merging galaxies as normal.} and merging galaxies. In our work we use the two\footnote{We tried also the relation based on the asymmetry and the Gini coefficient \citep{lotz2004}, but it gave a worse subdivision of the merging systems. Therefore, we rejected this relation.} main merger criteria:	

\begin{itemize}
\item \textbf{$A-S$ criterion} \citep{conselice2003}: in the plane $A$ vs. $S$ normal galaxies show a good correlation

\begin{equation}
	A_{\rm fit}(R) = (0.35 \pm 0.03) \times S(R) + (0.02 \pm 0.01) \, .
\label{eq:Afit}
\end{equation}
The two indexes are computed on $R$-band images, because they are less sensible to bright young stars and provide a more stable relation. Mergers should deviate from this relation because their light distribution, distorted from gravitational interactions, raises significantly the value of $A$, while it has a weaker influence on the $S$ parameter. Therefore, galaxies that 
show a large deviation from the fiducial sequence, or simply a very high value of asymmetry, that is
\begin{equation}
	A>A_{\rm fit}+3\sigma {\quad\rm or\quad} A>0.35
\label{eq:A-S}
\end{equation}
are classified as mergers ($\sigma$ is the mean dispersion in equation \ref{eq:Afit}, and is equal to 0.035).

\item \textbf{$G-M_{20}$ criterion} \citep{lotz2004}: as in the previous case, the correlation among normal galaxies in the plane $G$ vs. $M_{20}$ is used to define this merging criterion:
\begin{equation}
	G > -0.115 \times M_{20} + 0.384 \, .  
\label{eq:G-M}
\end{equation}
\end{itemize}

\subsection{Data processing} \label{sec:data-processing}
Our data processing workflow is organized into three main steps, each one coupled with a specific software.
\begin{itemize}
\item[-] Data acquisition: SDSS frames cover a field of view \mbox{of $\sim14\times10$ arcmin}. Because our galaxies are near and extended, they are often close to the edge of the image, or they fall halfway along multiple frames. We use the software \textsc{Montage}\footnote{ Developed by the NASA Earth Science Technology Office; http://montage.ipac.caltech.edu/.} to assemble multiple images in fits format (details about this step are given in Appendix \ref{app:montage}). \\
The Hyperleda database is an ideal starting catalogue for this operation since it provides, for each galaxy, the list of properties (coordinates, diameter, position angle, redshift, etc.) to automatically run \textsc{Montage}.
Because we are still dealing with a moderate number of sources, we carefully checked the correct assembly of all the images. 

\item[-] Pre-processing: in this step we prepare the image for the computation of the structural indexes: every feature that might affect the $CASGM$ analysis (i.e. bright stars in foreground, cosmic rays, image artifacts, etc.) must be masked. For our automated workflow, we used the software \textsc{SExtractor} \citep{bertin1996}, that provides a fast detection of all the sources in the image. Source identification is essentially based on local intensity and contrast, but the software examines also the light profile, extracting a number of properties (for a detailed description see Appendix \ref{app:sextractor}). 
Therefore, at the end of the pre-processing step\footnote{After the pre-processing phase, about 12 per cent of the sources is discarded, usually because the Hyperleda coordinates are wrong, or \textsc{Montage} can not produce the mosaic or it is impossible to setup the image properly (i.e. because the galaxy is too faint and is not fitted correctly by \textsc{SExtractor}).}, the original image is associated to a \textsc{SExtractor} catalogue, and to several \textquotedblleft service\textquotedblright{ }images, that specify the regions to exclude and provide useful information for the $CASGM$ analysis.

\item[-] $CASGM$ analysis: the crucial part of this work is entrusted to our software \textsc{PyCASSo} (Python $CAS$ Software), whose task  is to compute the $CASGM$ structural indexes. \textsc{PyCASSo} is entirely developed in \textsc{Python}, an high level and object oriented programming language, with extensive standard library and the possibility to import modules\footnote{In particular, \textsc{PyCASSo} needs \textsc{Numpy} scientific module, essential for matrix operations; \textsc{PyFITS}, used to read images in fits format; and \textsc{matplotlib}, used to create control images and plots.} suited for handling scientific data and astronomical images. 
We give a detailed description of the algorithms implemented in \textsc{PyCASSo} in Appendix \ref{app:pycasso}.
\end{itemize}

We tested our workflow and softwares on the \citet{frei} catalogue. This catalogue collects a sample of nearby, well-resolved galaxies, and it is therefore suitable for testing the reliability of the algorithm, possible side effects (see Appendix \ref{app:image-degradation} for an image degradation discussion) and improvements in the implementation of the $CASGM$ indexes. We compared our results on these galaxies with \citet{conselice2003}, \citet{lotz2004} and \citet{vikram2010} and we found a very good agreement: on average, the $C$, $A$, $S$ and $M_{20}$ coefficients are consistent within $1\sigma$ with the results of the other authors, while the Gini coefficient is in agreement within $1.5\sigma$. 
\\
\\
To further test the CASGM method, we carried out a visual classification on all the systems identified as merger by the CASGM analysis, both in the BAT and in the control samples. The visual classification assigns each galaxy to one of these three classes: (i) \textquotedblleft normal\textquotedblright{ }galaxies do not show any signs of interaction (i.e. appear regular and isolated); (ii) \textquotedblleft edge-on\textquotedblright{ }galaxies: these are intentionally kept separated from non edge-on galaxies to study possible biases related to dust bands, as highlighted by other studies (i.e. \citealp{jogee2009}; \citealp{depropris2007}); (iii) \textquotedblleft merger\textquotedblright{ }systems, i.e. close pairs of galaxies and sources showing morphological distortions or perturbations (such as tidal tails, double nuclei, etc.).
The visual classification is based first on the $RGB$ and fits images available in the SDSS database and on the corresponding fits images. Where available, we exploited also the spectroscopic data, to discern projected pairs of galaxies from real ones. Finally, for the most critical objects, we searched for further information in NED\footnote{NASA/IPAC Extragalactic Database,\\ http://ned.ipac.caltech.edu/.} and in the literature.

\section{Results}\label{sec:results}
Here we present the results of our automated classification and the merger fraction of the two samples. As explained in section \ref{sec:control-sample}, the BAT sample is a collection of systems, so we have to switch from galaxy to system classification also in the control sample, to make them fully comparable. To this purpose we consider as a single system any ensemble of galaxies for which one galaxy, at least, has been classified as interacting. The interacting or non-interacting classification is of course provided by the specific merger criterion considered. \\
We ran \textsc{PyCASSo} using both elliptical and circular apertures and we visually checked the control images and the results produced by our software. In most of the cases the two analyses coincide, but for some class of objects (i.e. edge-on galaxies and mergers) the elliptical apertures prove to be more reliable, being able to better fit the outline of these sources. In case of stretched objects, instead, circular apertures include a large amount of background, so the corrections applied to the asymmetry and the clumpiness become more critical. For these reasons, we report only the results of the elliptical classification. \\
\citet{lotz2004} studied the typical errors associated with the $CASGM$ measurements by analyzing the \citet{frei} images and the SDSS images of the same galaxy sample. These differences provide an average estimate of the uncertainties on the indices, in fact: (i) they take implicitly into account the slight smoothing effect introduced by \textsc{Montage} (because the \citet{frei} galaxies always belong to a single frame); (ii) they take into account the differences due to image resolution and quality (because \citet{frei} have a lower resolution, so they are similar to SDSS images at larger redshifts). The uncertainties related to the structural indexes are the following: $\delta C =0.11$, $\delta A =0.04$, $\delta S = 0.09$, $\delta G=0.02$, $\delta M_{20} = 0.12$.

\subsection{Results of the BAT sample}
\begin{figure}
\includegraphics[width=0.46\textwidth]{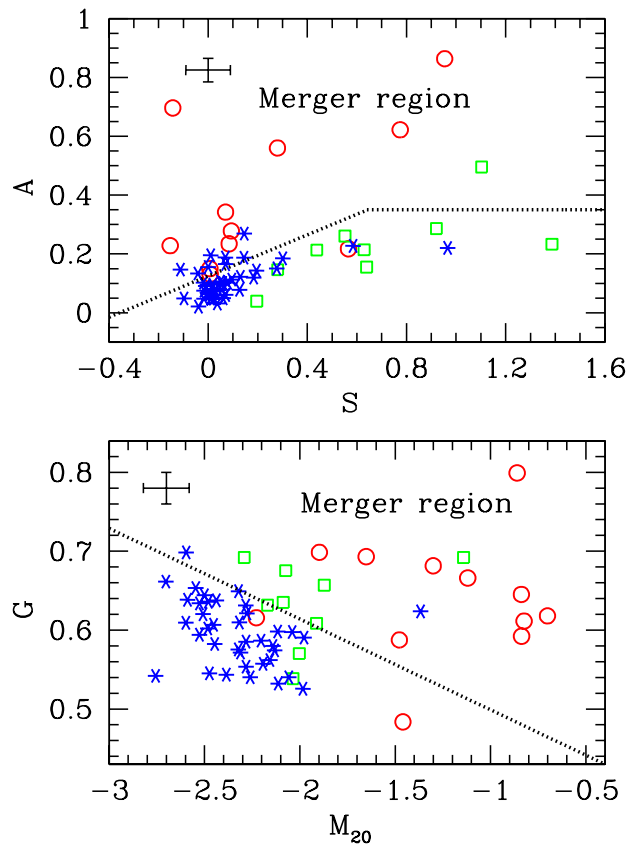}
\caption[]{Comparison between visual and automated classifications. Red circles: galaxies visually classified as interacting; green squares: edge-on galaxies; blue asterisks: normal galaxies. Structural indexes classify as merger the galaxies lying above the dotted lines. The error bars are average differences between SDSS and Frei observations of the same objects \citep{lotz2004}. The $A-S$ criterion shows a slight contamination produced by normal galaxies, while the $G-M_{20}$ is biased towards edge-on galaxies.}
\label{fig:bat-autom-class}
\end{figure}

The results obtained on the BAT sample, using both the visual and the automated classifications, are reported in \mbox{Table \ref{table:bat-classifications}} and discussed in the following (errors on the merger fractions are of 68 per cent confidence level and have been computed using the \citet{gehrels1986} prescriptions): 
\begin{itemize}
\item[-] Visual classification:  through the visual classification we estimate a merger fraction of $20_{-5}^{+7}$ per cent (we identify 12 mergers, 9 edge-on galaxies and 38 normal systems).
\item[-] $A-S$ classification: the criterion based on the asymmetry and the clumpiness (equation \ref{eq:A-S}) detects 20 mergers, giving a merger fraction of $34 \pm 7$ per cent. Eleven of the 12 systems visually classified as merger have the same classification with the $A-S$ method (see Table \ref{table:bat-classifications} and Figure \ref{fig:bat-autom-class}, upper panel).
The higher fraction of mergers detected with the $A-S$ method is due to a moderate contamination of normal systems with low clumpiness. 
In fact, for these cases, even a small asymmetry contribution, produced by small spurious\footnote{For example, in some cases \textsc{SExtractor} is not able to separate the faint high redshift galaxies in the background from the main one, and the same occurs for small stars in the foreground. If the clumpiness value is near zero, the asymmetry contribution coming from these sources may determine their misclassification as mergers.} sources within the $CAS$ aperture, may be enough for labelling that galaxy as interacting.
\item[-] $G-M_{20}$ classification: the criterion based on the Gini coefficient and the momentum of light (equation \ref{eq:G-M}) identifies 18 mergers, giving a merger fraction of $31\pm7$ per cent. In this case, the higher fraction of merging systems with respect to the visual classification is due to the contamination produced by edge-on galaxies. These galaxies are observed through dust bands that obscure the central part of the source and leave two bright areas symmetrically off-centred that influence the momentum of light. A similar effect occurs also for pronounced barred galaxies. Ten of the 12 systems visually classified as mergers have the same classification also through the $G-M_{20}$ method (see Table \ref{table:bat-classifications} and Figure \ref{fig:bat-autom-class}, lower panel). 
\end{itemize}
\subsubsection{Improvement of the $CASGM$ system}\label{sec:casgm-improvement}
As shown in the previous section, the automated classifications correctly identify almost all the interacting systems, but they systematically overestimate the real number of mergers. For this reason, we introduce an advanced criterion that blends together\footnote{Some attempts have been made by \citet{lotz2004}, that studied other combinations of $CAS$ and $GM$ indexes (i.e. \mbox{$G-A$}, \mbox{$G-S$}, \mbox{$A-M_{20}$}) which, however, did not produce better classifications than the $A-S$ and $G-M_{20}$ criteria.}  the previous procedures: we consider as mergers only those systems that satisfy simultaneously the $A-S$ and the $G-M_{20}$ criteria. All these indexes have similar resolution and $S/N$ requirements and so they can be used together; however, this choice may limit the effectiveness of the merger identification, because each method is not sensible to the entire duration of the merger and the interaction phases mapped by each criterion do not fully overlap (see \citealp{lotz2008b} and \citealp{conselice2006}). 
We expect the combined criterion to be much more reliable than the original ones. For instance, the $G-M_{20}$ contamination should be largely removed because edge-on and barred galaxies are basically symmetric and, therefore, they should be excluded by adding the $A-S$ classification.

\subsubsection{Merger fraction of the BAT sample}
\begin{table}
\centering
\begin{tabular}{c | c | c c c}
\textbf{PBCJ} & \textbf{Visual} & \multicolumn{3}{c|}{\textbf{Automated classification}} \\
\textbf{name} &	\textbf{analysis} &	\textbf{$A$-$S$} & \textbf{$G$-$M_{20}$} &	Combined crit. \\
\hline
0042.8-2331	&	$M$	&	$\times$	&	$\times$	&	$\times$	\\
0124.4+3346	&	$M$	&						&						&						\\
0209.4-1010	&	$M$	&	$\times$	&	$\times$	&	$\times$	\\
0241.5+0709	&	$M$	&	$\times$	&						& 					\\
0252.4-0832	&	$e$	&						&	$\times$	&						\\
0255.2-0011	&	$M$	&	$\times$	&	$\times$	&	$\times$	\\
0303.8-0107	&	$n$	&	$\times$	&						&						\\
0742.4+4498	&	$n$	&	$\times$	&						&						\\
0744.1+2915	&	$M$	&	$\times$	&	$\times$	&	$\times$	\\
0759.9+2324	&	$n$	&						&						&						\\
0823.0-0454	&	$M$	&	$\times$	&	$\times$	&	$\times$	\\
0919.9+3712	&	$e$	&						&	$\times$	&						\\
0926.1+1245	&	$n$	&	$\times$	&						&						\\
0942.1+2342	&	$n$	&	$\times$	&						&						\\
1002.0+5539	&	$e$	&	$\times$	&	$\times$	&	$\times$	\\
1023.5+1951	&	$M$	&	$\times$	&	$\times$	&	$\times$ 	\\
1104.4+3813	&	$M$	&	$\times$	&	$\times$	&	$\times$ 	\\
1113.7+0930	&	$n$	&	$\times$	&						&						\\
1139.6+3157	&	$M$	&	$\times$	&	$\times$	&	$\times$	\\
1204.4+2018	&	$n$	&	$\times$	&						&						\\
1206.2+5244	&	$n$	&	$\times$  &	$\times$	&	$\times$	\\
1217.1+0712	&	$e$	&						&	$\times$	&						\\
1225.7+1240	&	$e$	&						&	$\times$	&						\\
1345.4+4141	&	$e$	&						&	$\times$	&						\\
1417.9+2508	&	$n$	&	$\times$	&						&						\\
1424.3+2436	&	$n$	&						&	$\times$	&						\\
2236.0+3358	&	$M$	&	$\times$	&	$\times$	&	$\times$ 	\\
2318.9+0014	&	$M$	&	$\times$	&	$\times$	&	$\times$ 	\\
\hline
Total mergers	&	12		&	20		&	18				&	12				\\
\hline
\end{tabular}
\caption{Merger identifications in the BAT sample: we report only those galaxies that have been tagged as interacting by at least one classification method. In the visual classification, \textquotedblleft $M$\textquotedblright{ }identifies mergers, \textquotedblleft $n$\textquotedblright{ }the non interacting galaxies and \textquotedblleft $e$\textquotedblright{ }the edge-on galaxies. The mergers of the automated criteria are labeled by a \textquotedblleft $\times$\textquotedblright{ }mark. It is possible to notice that the combined criterion is much more reliable than the others, in fact it removes most of the contaminations and it provides results in good agreement with our visual analysis.}
\label{table:bat-classifications}
\end{table}
\begin{figure}
\includegraphics[width=0.46\textwidth]{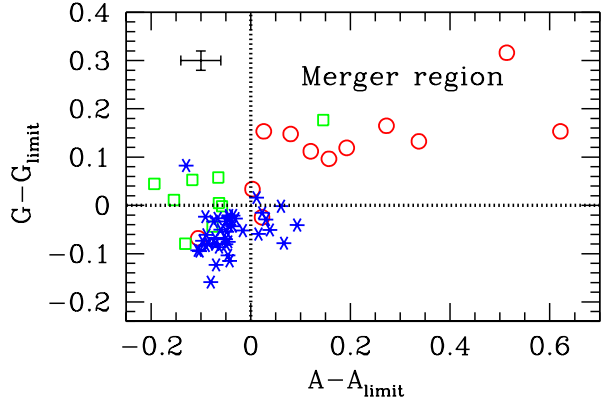}
\caption[]{Comparison between our combined criterion and the visual classification: the structural indexes classify as merger the galaxies lying in the top right-hand sector, while symbols and colors are the same as in Figure \ref{fig:bat-autom-class}. The combined criterion shows a good agreement with our classification, in fact the contaminations affecting the original criteria are almost completely removed.}
\label{fig:bat-asgm}
\end{figure}
The combined criterion proves to be an optimal solution, in fact it does not miss almost any merger compared to the previous criteria and it removes about 77 per cent of their wrong classifications, leading to a merger identification in excellent agreement with our visual analysis (see Table \ref{table:bat-classifications} and Figure \ref{fig:bat-asgm}). By exploiting the combined criterion, we detect 12 disturbed systems, so the merger fraction of the BAT sample is $20_{-5}^{+7}$ per cent. \\
Even if the low statistics does not allow any strong conclusions, we point that the merger fraction among each redshift bin is almost constant (Table \ref{table:result-z-bin}), so it does not display any evident signs of evolution in the local Universe. \\
Our results are in  excellent agreement with \citet{koss2010}, that performed a visual analysis on a similar BAT subsample and found a merger fraction of 25 per cent, by considering all the perturbed galaxies and the pairs with a separation below 30 kpc. 
We have compared the luminosity distributions (14-150 keV band) of the interacting and the non interacting systems of our BAT sample and, according to a KS test ($\rm {prob}_{\rm ks} =0.2$), the luminosity distributions of the two subsamples do not display significant differences. 
\begin{table*}
\begin{minipage}{\textwidth}
\centering
\begin{tabular}{c | c c l | c c l}
\textbf{Redshift} & \multicolumn{3}{c|}{\textbf{BAT sample}} & \multicolumn{3}{c|}{\textbf{Control sample}} \\
 & \textbf{$N_{\rm sys}$} &	\textbf{$N_{\rm m}$} &	\textbf{$\quad f_{\rm m}$}  & \textbf{$N_{\rm sys}$} &	\textbf{$N_{\rm m}$} &	\textbf{$\quad \,\,\, f_{\rm m}$} \\
\hline
$0.003 \leq z < 0.01$ & 15 & 3 & 20  (9-36) & 63  & 3 & 4.8 (2.2-9.2) \\
$0.01  \leq z < 0.02$ & 16 & 3 & 19  (9-34) & 67  & 4 & 6.0 (3.1-10.5) \\
$0.02  \leq z < 0.03$ & 28 & 6 & 21 (13-32) & 117 & 9 & 7.7 (5.2-11.0) \\
\hline
Total ($CASGM$) & 59 & 12 & 20 (15-27) & 247 & 16 & 6.5 (4.9-8.5) \\
Corrected & 59 & 12 & 20 (15-27) & 247 & 10 & 4.0 (2.8-5.7) \\
\hline
\end{tabular}
\caption{Detailed comparison of the merger fraction $f_{\rm m}$ of the BAT and of the control sample in each redshift bin, according to the classification of the combined criterion. In the \textquotedblleft Total $CASGM$\textquotedblright{ }line are summarized the results of the mere $CASGM$ classification, while in the \textquotedblleft Corrected\textquotedblright{ }line we indicate the merger fractions after the application of the reliability and the completeness corrections.
AGN host galaxies are found more frequently in phase of interaction compared to a random selection of galaxies in the same redshift interval. This suggests that there is a link between the merging event and the activity of the SMBH at the centre of galaxies.}
\label{table:result-z-bin}
\end{minipage}
\end{table*}

\subsubsection{Statistical corrections}\label{sec:stat-correction}
It is possible to further improve our results, by estimating the completeness and the reliability of the automated classification and applying a statistical correction to the merger fraction.
\begin{itemize}
\item[\textbf{-}] \textbf{Completeness:} it quantifies the amount of missed mergers, that is the number of systems that have been labelled as \textquotedblleft interacting\textquotedblright{ }by the visual classification, but as \textquotedblleft non interacting\textquotedblright{ }by the combined criterion. We define this coefficient as
\begin{equation}
	C_{CASGM} = \frac{N_{\rm m,true}}{N_{\rm m,visual}} \, ,
	\label{eq:completeness}
\end{equation}
where $N_{\rm m,true}$ is the number of mergers in common between the automated and the visual classification, while $N_{\rm m,visual}$ represents the number of mergers of the visual classification. By extrapolating the completeness from the BAT sample, we obtain $C_{CASGM} = 10/12 = 0.8_{-0.2}^{+0.1} \,$. This parameter allows to derive the real merger fraction of the sample, in fact it tells that the number of mergers that have been correctly\footnote{Spurious and wrong merger detections must be excluded from the sum.} detected by the automated classification is about $80$ per cent of the real number.
\item[\textbf{-}] \textbf{Reliability:} it quantifies the fraction of normal systems that have been erroneously classified as mergers by the automated procedure. We define it through the probability, $P$, that the procedure gives a false positive (false merger) in case of a non-merging system, i.e.:
\begin{equation}
P_{CASGM} = \frac{N_{\rm m,false}}{N_{\rm normal}} \, ,
\end{equation}
where $N_{\rm m,false}$ is the number of wrong mergers and $N_{\rm normal}$ is the number of non interacting sources (that is the difference between the number of systems $N_{\rm sys}$ in the sample and the number of real mergers $N_{\rm m,real}$). By extrapolating this value from the BAT sample, we obtain: $P_{CASGM} = 2/47 \sim 0.04_{-0.03}^{+0.06} \,$, which means that about 4 per cent of the non interacting systems is instead classified as merger by the combined criterion.
\end{itemize}

A good knowledge of these coefficients is extremely useful for correcting the merger fraction of very large samples, that can not be visually inspected. In fact, by applying the reliability correction, we obtain the number of \textquotedblleft true\textquotedblright{ }mergers detected by the software, and then, taking into account the completeness coefficient, we can estimate the real number of interacting systems $N_{\rm m,real}$:
\begin{equation}
	N_{\rm m,real} = \frac{N_{\rm m}-P_{CASGM}\times N_{\rm sys}}{C_{CASGM}-P_{CASGM}} \, ,
\label{eq:N_m_real}
\end{equation}
where $N_{\rm m}$ is the number of mergers detected by the combined criterion and $N_{\rm sys}$ is the total number of systems in the sample.

\subsection{Results of the control sample}

\subsubsection{Merging fraction and statistical corrections}\label{sec:control-sample-corrections}
The procedure described in the previous sections detects 16 merging systems in the control sample (see Table \ref{table:result-z-bin}) corresponding to a merger fraction of $f_{\rm m,control}=6.5_{-1.6}^{+3.0}$ per cent. This fraction, however, does not take into account the corrections for the reliability and the completeness previously discussed. 
Using our estimates of $P_{CASGM}$ and $C_{CASGM}$ based on the BAT sample, we derive that the real number of mergers in the control sample is (see equation \ref{eq:N_m_real}) $N_{\rm m,real} \sim 8$.
In particular, the expected number of true mergers among the 16 detected by the algorithm is $\sim 6$ ($P_{CASGM}$ correction), while two more real mergers are expected to be missed by the procedure ($C_{CASGM}$ correction).
Given the large fraction of the detected mergers that are expected to be spurious (more than 60 per cent), we have visually inspected  all the 16 systems found by the procedure as mergers, to confirm and better constrain the actual number of false/true mergers. In good agreement with our expectations, we find that only 8 systems are true mergers, the remaining ones being star-burst or irregular galaxies. This number confirms that the procedure works similarly in the  BAT and in the control sample. 

 By applying also the completeness correction
  (equation \ref{eq:completeness}), we derive that the total number of
  real mergers in the control sample is 8/0.8$\sim$10 which corresponds
  to a merger fraction of: 
\begin{equation}
f_{\rm m,corr,control} \simeq 10/247 \simeq 4.0_{-1.2}^{+1.7} \,\,\,{\rm
  per \,\, cent.}
\end{equation}

In addition, the large number of objects in this sample allows us to derive an estimate of $P_{CASGM}$ which is more accurate than the one based on the BAT sample:
\begin{equation}
	P_{CASGM} = (8_{-2.77}^{+3.95})/237 \simeq 0.034_{-0.012}^{+0.017} \, .
\end{equation}

Our results show that the average merger fraction of galaxies at redshift $\sim 0$ is very low, in accordance with the studies of \citet{patton2008}, \citet{patton2002} and \citet{koss2010}, that claim a merger fraction of $\sim 1-2$ per cent. The higher value suggested by our work is probably related to a selection effect, because our control sample is not drawn as a random selection of galaxies in the prefixed redshift interval, but it is forced to follow the BAT sample's redshift distribution. This confirms the importance of building a control sample which reflects, as much as possible, all the key properties of the other sample. The merger fraction found in the control sample is significantly (3$\sigma$) lower than that found in the BAT sample.

\begin{figure}
\includegraphics[width=0.46\textwidth]{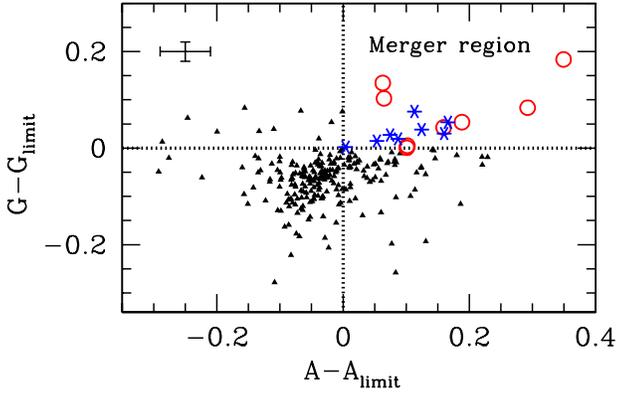}
\caption[]{Classification of control sample's systems according to the combined criterion: the systems lying in the top right-hand sector of the plot are labelled as mergers by the automated criterion. For these objects, we performed a visual analysis (red circles: galaxies visually classified as interacting; blue asterisk: normal galaxies), while black triangles represents non-merger systems according to $CASGM$.}
\label{fig:control-asgm}
\end{figure}

\subsubsection{The role of the galaxy mass distribution}\label{MD-effects}
The (stellar) mass distribution of galaxies hosting BAT AGNs is very
likely to be different from that of inactive galaxies or SDSS AGNs
(Koss et al. 2011), with BAT AGNs typically residing in galaxies more
massive than average.

The effects of galaxy mass upon the merger fraction measured through
the $CASGM$ method are uncertain\footnote{\citet{patton2008} find that
  the frequency of galaxy pairs is larger for low-luminosity (and,
  presumably, low-mass) than for high-luminosity galaxies; but this
  trend is reversed when they correct for perspective pairs. The $CASGM$
  method is somewhat in between the two cases: a galaxy pair which is
  well-separated on the sky will be classified as a merger only if
  there are morphological anomalies (i.e. if the pair is physical);
  but the method cannot distinguish physical and perspective pairs if
  the sky separation is small. Then, the $CASGM$-measured merger
  fraction should have only a weak dependence on galaxy
  mass.}. However, if the mass dependence is relatively strong, we
might obtain different merger fractions for the BAT and the control sample
simply because of their different mass distributions. Therefore, it is
necessary to check this hypothesis.

As a first step we evaluated galaxy stellar masses: this was done by
converting the $ugriz$ magnitudes from the SDSS into Johnson $BVRI$
magnitudes (using the formulae in \citealt{blantonroweis2007}),
calculating the distance modulus (DM) from the redshift of each galaxy
(we assumed $H_0=71\; {\rm km\, s^{-1}\, Mpc^{-1}}$), and finally
estimating the stellar masses as
\begin{equation}
\log{(M_*/{\rm M_\odot})} =
\log{[\mathcal{M}_{\rm I}(B-R)]} + 0.4(I-DM-I_\odot),
\end{equation}
where $\mathcal{M}_{\rm I}(B-R)$ is the mass to light ratio (in solar units)
provided by \cite{belldejong2001} for the $I$ band, and as a function
of the $B-R$ colour of the galaxy; whereas $I$ is the galaxy apparent
$I$ magnitude, and $I_\odot = 4.52$ is the $I$ absolute magnitude of
the Sun.

\begin{figure}
\includegraphics[width=0.46\textwidth]{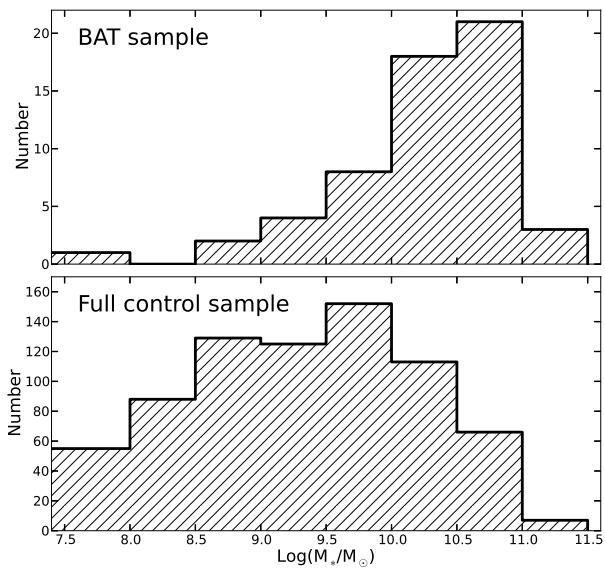}
\caption[]{Comparison between the mass distributions of the BAT sample
  (top panel) and of the full control sample (bottom panel).}
\label{fig:Mfunc-comparison}
\end{figure}

Figure \ref{fig:Mfunc-comparison} compares the distributions of
stellar masses in the BAT and in the full control sample: the two
distributions are quite different, as massive galaxies are much more
frequent in the BAT sample. We note that this is partly caused by the
contribution of the AGNs within the galaxies of the BAT sample;
however, the observed difference in luminosities is very large (the
medians of the two samples differ by a factor of $\sim 5$), and cannot
be explained in this way.

We checked whether this difference in the mass distributions could
account for the difference in the merger fraction by building a
mass-matched sample in the same way as we built a redshift-matched
sample (see Sec. 2.2). In this case, we divided the galaxies in 3 mass
bins ($M_*/{\rm M_\odot} < 10^{9.5}$; $10^{9.5}\le M_*/{\rm M_\odot} <
10^{10.5}$; $M_*/{\rm M_\odot} \ge 10^{10.5}$), and extracted 173
systems from the full control sample.

Within the mass-matched control sample, 11 systems are classified as
mergers by the $CASGM$ combined criterion; this corresponds to an uncorrected
merger fraction $f_{\rm m,MMS} = 11_{-3.3}^{+4.4}/173 \simeq
6.4_{-1.9}^{+2.5}$ per cent, and to a corrected merger fraction of
$f_{\rm m,corr,MMS}=3.9_{-2.4}^{+3.3}$ per cent, in very good
agreement with the values for the redshift-matched control sample.

 This result should be taken with caution, since the redshift
  distribution of the mass-matched control sample is different from
  that of the BAT sample. An ideal comparison should use a sample that
  {\it simultaneously} matches both the mass and redshift
  distributions of the BAT sample; unfortunately, our full control
  sample does not allow to proceed in this way, as it includes only a
  small number (5) high-mass ($M_*/{\rm M_\odot} \ge 10^{10.5}$)
  systems at $z < 0.02$.

 However, we can look at the simultaneous effect of both mass and
  redshift in two different ways. In the $0.02\le z < 0.03$ bin the
  full control sample includes a reasonable number (68) of high-mass
  systems: therefore, we extracted a mass-matched control sample
  within this redshift bin, where the combined criterion finds 15
  mergers among the 162 systems. This corresponds to an uncorrected
  merger fraction $f_{\rm m,MMS,z \ge 0.02} = 15_{-3.8}^{+5.0}/162 \simeq
  9.3_{-2.3}^{+3.1}$ per cent, and to a corrected merger fraction of
  $f_{\rm m,corr,MMS,z \ge 0.02}=6.3_{-3.0}^{+3.9}$ per cent; both values
  are consistent with the results for the same redshift bin that we
  gave in Table 2 ($f_{\rm m,z \ge 0.02}=7.7_{-2.5}^{+3.3}$ per cent, and
  $f_{\rm m,corr,z \ge 0.02}=5.6_{-3.2}^{+4.6}$ per cent).

Instead, when looking at our full redshift range,
we evaluate the uncorrected merger fraction in
each bin of redshift and mass, and average them so as to reproduce the
mass and redshift distribution of the BAT sample. In this way, we get
an uncorrected merger fraction $f_{\rm m,avg} = 7.2_{-2.7}^{+9.1}$ per
cent, and a corrected merger fraction\footnote{If $f_{\rm m} \equiv
  N_{\rm m}/N_{\rm sys}$ is the uncorrected merger fraction,
  equation \ref{eq:N_m_real} implies that $f_{\rm m,corr} \equiv N_{\rm
    m,real}/N_{\rm sys} = (f_{\rm m}-P_{\rm CASGM})/(C_{\rm CASGM}-P_{\rm
    CASGM})$.}  $f_{\rm m,avg,corr}=3.7_{-3.4}^{+11.5}$ per cent. The
large errors derive from the highly uncertain merger
fractions of high-mass systems at $z<0.02$: if instead we make the
very reasonable assumption that these are equal to what we find for
high-mass systems at $0.02 \le z < 0.03$ ($f_{\rm
  m,z\ge0.02,log(M) \ge 10.5}=8.8_{-3.5}^{+5.3}$ per cent, fully
compatible both with the scarce high-mass data at $z<0.02$, and with
the redshift trend of the merger fractions in the other mass bins), we
obtain an uncorrected merger fraction $f_{\rm m,avg*}=5.9_{-1.6}^{+3.5}$ per cent,
and a corrected one of $f_{\rm m,avg*,corr}=2.5_{-2.1}^{+4.6}$ per cent.

We conclude that simultaneously controlling for the mass and
  redshift distributions cannot reconcile the merger fractions of the
  BAT and the control sample. This fact is proved (at the $1.8\sigma$
  level) for the $0.02 \le z < 0.03$ redshift bin. In the full
  sample it somewhat depends on the assumption that the merger
  fraction for galaxies with $M_* \ge 10^{10.5}\, {\rm M_\odot}$ does
  not change between $z=0.003$ and $z=0.03$: if such assumption is
  made, the (corrected) merger fractions of the two samples differ at
  the $2.6\sigma$ level.

\section{Summary and conclusions} \label{sec:conclusions}
In this work we focused on three main topics:

(i) Software: we have implemented the new software \textsc{PyCASSo} for the automated computation of the structural indexes of the $CASGM$ system. Our procedures are entirely based on the definitions and relations presented in \citet{conselice2003} and \citet{lotz2004}, but we have implemented the possibility to use elliptical apertures, because they provide a better fit of the galaxy outline. Moreover, we carried on extensive tests on possible image degradations, so our software minimizes any data loss and smoothing effect and provides a stable and reliable analysis. \\

(ii) Method: we propose an improved technique for evaluating the merger fraction of a galaxy sample by means of the $CASGM$ system. Indeed, we show that the original classification is biased towards irregular, edge-on and dusty galaxies, which tend to be misclassified as mergers. We propose a combined criterion between the $A$, $S$, $G$ and $M_{20}$ indexes, which leads to the complete blending of the $CAS$ and $GM$ methods and corrects nearly 80 per cent of the contamination. Then, we define the completeness and the reliability coefficients, that allow a statistical correction of the merger fraction and further reduce possible residual errors in the automated classification. \\

(iii) Application: we have applied the $CASGM$ analysis to a sample of
local AGN host galaxies and a comparison sample, to extract their
merger fractions and test whether there is an enhanced fraction of
mergers among active galaxies.  We found that in the BAT sample the
merger fraction is $20_{-5}^{+7}$ per cent.  In the
  redshift-matched control sample the merger fraction is
  $4.0_{-1.2}^{+1.7}$ per cent, and the difference is significant at
  the $3\sigma$ level. We obtain similar results for a mass-matched
  control sample. Simultaneously matching redshift and mass leads to
  comparable but somewhat less significant results.

Our work is in agreement with other observational studies
(\citealp{sanders1988}; \citealp{koss2010}) and numerical simulations
(\citealp{barnes1991}; \citealp{dimatteo2005}; \citealp{hopkins2006})
that suggest that galaxy interactions trigger the activity of the SMBH
at their centre. The most likely scenario is that the strong
gravitational perturbations drive large quantities of gas towards the
centre of the remnant, originating both an intense starburst phase and
an enhanced nuclear activity. Mergers may therefore be responsible not
only for large scale ($\sim 10^3$ pc) distortions, but also of the
inflow of gas down to the typical scale of SMBH accretion ($\sim
10^{-4}$ pc). Current numerical simulations can not investigate
entirely such a wide scale range, so observational studies have a key
role for the comprehension of these phenomena.
However, as we pointed out in Section \ref{sec:intro}, similar
  studies on higher redshift ($0.2<z<1.2$) galaxy samples
  (i.e. \citealp{cisternas2011}; \citealp{gabor2009};
  \citealp{pierce2007}) do not show any enhancements of the merger
  fraction of AGN host galaxies. Selection biases in the active
  galaxies sample and/or in the control sample could partially explain
  these contradicting results. For example, the aforementioned studies
  are based on other selection criteria (i.e. soft X-ray, 2--10 keV
  energy band), but, due to the significant fraction of obscured AGNs
  (see Menci et al.  2008), they may detect a lower number of sources
  compared to our hard X-ray (15-195 keV) selection.

 Therefore, while the results presented here and in previous
  observational studies (e.g. \citealp{koss2010}) suggest that in the
  low redshift ($z < 0.03$) Universe galaxy interactions trigger the
  activity of the SMBH at their centre, further researches that focus
  on an accurate and unbiased selection of galaxies - both at
  intermediate ($0.03 \le z < 0.2$) and higher ($0.2 \le z < 1.2$) redshifts -
  are mandatory to derive improved estimates on the occurrence and
  role of galaxy interactions on SMBH activity.

\section*{Acknowledgments}
We gratefully acknowledge V. Vikram for his access to his morphology analysis code and Valentina La Parola for a careful reading of the manuscript and for her useful comments. 
We thank the anonymous referee for her/his detailed comments that have improved the quality of the paper.   
We thank CILEA Consortium for the giving us access to the HPC cluster Lagrange. This research made use of Montage, funded by the National Aeronautics and Space Administration's Earth Science Technology Office, Computation Technologies Project, under Cooperative Agreement Number NCC5-626 between NASA and the California Institute of Technology. Montage is maintained by the NASA/IPAC Infrared Science Archive.\\
Funding for SDSS-III has been provided by the Alfred P. Sloan Foundation, the Participating Institutions, the National Science Foundation, and the U.S. Department of Energy Office of Science. The SDSS-III web site is http://www.sdss3.org/.
SDSS-III is managed by the Astrophysical Research Consortium for the Participating Institutions of the SDSS-III Collaboration including the University of Arizona, the Brazilian Participation Group, Brookhaven National Laboratory, University of Cambridge, Carnegie Mellon University, University of Florida, the French Participation Group, the German Participation Group, Harvard University, the Instituto de Astrofisica de Canarias, the Michigan State/Notre Dame/JINA Participation Group, Johns Hopkins University, Lawrence Berkeley National Laboratory, Max Planck Institute for Astrophysics, New Mexico State University, New York University, Ohio State University, Pennsylvania State University, University of Portsmouth, Princeton University, the Spanish Participation Group, University of Tokyo, University of Utah, Vanderbilt University, University of Virginia, University of Washington, and Yale University.\\
We acknowledge the usage of the HyperLeda database (http://leda.univ-lyon1.fr). The authors acknowledge partial financial support from ASI (grant n. I/088/06/0, COFIS contract and grant n. I/009/10/0).

\appendix

\section[]{Data processing algorithms}\label{app:data-processing}

The image analysis process is split into three main phases: data acquisition, pre-processing and processing. In the first two phases we essentially use publically available codes (\textsc{Montage} and \textsc{SExtractor}), whereas for the processing phase we developed the software \textsc{PyCASSo}. 

\subsection{Montage: data acquisition}\label{app:montage}
We use the software \textsc{Montage} to automatically assemble multiple SDSS frames, in order to obtain full images of the desired galaxies.
This software needs as input the central coordinates, the band of observation and the sizes (arcmin) of the desired field of view. It automatically queries the SDSS database for the frames that compose the image, and, exploiting the astrometry and the calibrations of the original frames, it proceeds with their alignment and superposition, it compares the intensities of the overlapping pixels and it corrects possible background offsets and gradients, to produce a uniform mosaic. \textsc{Montage} preserves all the information of the original images (such as the photometric intensity of the sources and the World Coordinate System), it is able to assemble together a large number of frames and has a very good success rate (more than 95 per cent in our experience) so it is the ideal instrument for our automated workflow. Moreover, the images returned by \textsc{Montage} are centred on the selected coordinates, so the queried galaxy is always in the middle of the frame.\\
\textsc{Montage} has two drawbacks: Moir\'e pattern and slight image degradation. The first is an interference pattern that occurs when two grids, with different orientations or mesh sizes, are superimposed. This is unavoidable, because \textsc{Montage} has to create a new grid of pixels (the final image) and overlap portions of the original frames, rotating them properly to match each other. \citet{conselice2003} carried on numerical simulations to determine the impact of correlated noise on the asymmetry measure: it turned out that this effect is very small ($\delta A\leq0.03$ on average) because the $CASGM$ background correction routine, by analysing a region of pure background, takes into account the noise pattern. Moreover, in our case, the Moir\'e pattern appears only by giving much contrast to the images, so it is generally unimportant. \\ 
Image degradation (see Appendix \ref{app:image-degradation}) inevitably occurs because, during a rotation or a non-integer translation, each new pixel is the average of the original pixels which lay in that position, each one weighed on its fraction of area. 
It is almost impossible to assess the amount of smoothing caused by image degradation, because it depends on the number of frames assembled, and on how they overlap. Our numerical tests suggest that this effect can reduce the background variance by a factor between $\sim{}$ 1 and 4; this estimate is confirmed by our measures on BAT galaxy images, that give an average reduction factor of 1.5. The plus side is that, again, background corrections reduce this error, because the smoothing affects both galaxy and background light distribution.

\subsection{SExtractor: pre-processing}\label{app:sextractor}
We exploit \textsc{SExtractor} during the pre-processing step, in order to detect all the sources inside the image and identify those that that need to be masked because they may alter the $CASGM$ analysis (i.e. bright stars or cosmic rays). The software first examines the light profile of each source, and then extracts a catalogue of properties related to their photometry. It is also possible to save control images, such as the SEGMENTATION map (where all the pixels belonging to the same source have the same value, corresponding to the source ID number reported in the catalogue). To remove undesired sources, we exploit the CLASS\_STAR parameter returned by \textsc{SExtractor}: this specifies whether the light profile of the source is point-like/stellar (CLASS\_STAR$\sim 1$) or extended/non-stellar (CLASS\_STAR $\sim 0$). Our pre-processing uses a simple script that checks CLASS\_STAR values: if it is greater that 0.1, it flags as \textquotedblleft star\textquotedblright{ }the corresponding line of the catalogue. In the processing phase, all the stellar sources will be masked, according to the outline provided by the SEGMENTATION image. Of course, the automated procedure is efficient, but not always perfect: the script warns the user in case of conflicting results and it is possible to edit the mask and add custom circular or elliptical masks. 
Since we want to evaluate the merger fraction we do not manually remove any chance superpositions. The \textquotedblleft normal\textquotedblright{ }or \textquotedblleft merger\textquotedblright{ }classification is uniquely provided by the CASGM analysis. 

\subsection{PyCASSo: $CASGM$ analysis}\label{app:pycasso}
\textsc{PyCASSo} is in charge of the core of our analysis, that is the computation of the structural indexes for the automated classification of galaxies. Our software is entirely written in \textsc{Python}, making use of standard extension modules (e.g. \textsc{Numpy}, \textsc{PyFITS}, etc.), and it can be run both interactively and in batch mode, providing a fast analysis for each galaxy\footnote{For example, \textsc{PyCASSo} takes $\sim100$ sec to analyse a $900 \times 900$ image, when running on an Intel Celeron CPU at 2.0Ghz, with 2GB RAM.}. 
\\

Here we give a concise description of its workflow, while in the following sections we describe in detail the features of the software and their implementation. In the development process we paid particular attention to possible image degradation effects, so we will point out also some differences between our implementation and those of other authors (\citealp{conselice2003}, \citealp{vikram2010}).
\\
\textsc{PyCASSo} loads all the data computed in the previous steps, masks the unwanted sources, and subtracts the image background, providing a \textquotedblleft clean\textquotedblright{ }image. Then, it selects the target galaxy and computes, through a recursive process, its position angle, petrosian radius and asymmetry index\footnote{For the $CAS$ indexes, the centre of the galaxy is the pixel which minimizes the asymmetry value, so the asymmetry must be recomputed after each variation in the estimate either of the position angle and of the petrosian radius of the galaxy.}, and extracts - according to the definition given in Section \ref{sec:casgm-parameters} - the aperture that defines the area of the source (Appendix \ref{app:pycasso-asymmetry}). Using the same aperture, it computes the concentration (Appendix \ref{app:pycasso-concentration}) and the clumpiness (Appendix \ref{app:pycasso-clumpiness}). 
In case of pairs of galaxies, the companion can be included (partially or entirely) in the aperture: in this case it is necessary that the two galaxies are close enough,\footnote{It is unlikely to detect pairs of galaxies with separation greater than 30 kpc as mergers. In fact, the $CAS$ aperture can not extend much beyond the edge of the first galaxy and in general the whole $CASGM$ system is sensible only to pairs of galaxies close enough to perturb each other.} and that they are somehow connected each other (i.e. by a tidal tail or by a luminous halo), because the cut-off of the aperture is based on the light profile of the first galaxy, so it can not extend much beyond its outline. If there is a clear separation between the light distributions of the two sources, the aperture fits tightly the first galaxy and the companion is automatically excluded.
Next, \textsc{PyCASSo} corrects the estimates of asymmetry and clumpiness for any possible contributions coming from the background (Appendix \ref{app:pycasso-corrections}): this step is crucial, in fact the correction can lead also to the halving of the original values. At this point the software picks up again the \textquotedblleft clean\textquotedblright{ }image and extracts the segmentation map of the galaxy (see Appendix \ref{app:pycasso-gini-momentum}): the map follows the galaxy contour and again, in case of close pairs, it includes both the galaxies. On the contrary, all the sources that satisfy the brightness constraints of the segmentation map, but are not directly linked with the main galaxy, are masked. The segmentation map is then used by \textsc{PyCASSo} for computing the Gini coefficient and the second order momentum of light according to their definitions. 
The $CASGM$ indexes and all the other parameters that are computed by \textsc{PyCASSo} are collected in an ascii file. The software saves also a set of control images and warns the user if the galaxy size is too small to allow a reliable analysis.

\subsubsection[]{Image preparation}\label{app:pycasso-image-preparation}
\textsc{PyCASSo} extracts the positions and properties (semi-major axis, position angle, axis ratio) of the galaxy to analyse from \textsc{SExtractor} catalogues generated in the pre-processing phase. Such catalogues (and the associated SEGMENTATION image, see Section \ref{app:sextractor}) are used to mask out contaminating sources. The intensity of the background is evaluated as the mode (calculated as $3\times median - 2\times average$, cfr. \citealt{kendall1977}) of the pixels surviving a recursive sigma-clipping algorithm. Such intensity is subtracted from the masked image, and the result is used in all the following steps of the $CASGM$ analysis.

\subsubsection[]{Properties of the main galaxy}\label{app:pycasso-galaxy-properties}
The $CAS$ method is usually applied on a circular area with radius $1.5r_{\rm P}$ ($r_{\rm P}$ is the Petrosian radius, see Section \ref{sec:casgm-parameters}). However, \textsc{PyCASSo} can also use elliptical areas, because they are often more suited for stretched galaxies and close pairs (i.e. the kind of objects we are most interested in).
When using elliptical areas, we consider $r_{\rm P}$ as the ellipse's semi-major axis, and we use the axis ratio in the \textsc{SExtractor}  catalogues. Since the position angle in the \textsc{SExtractor} catalogue is often inaccurate, we recompute it (by maximizing the flux inside the elliptical area).

\subsubsection[]{Asymmetry}\label{app:pycasso-asymmetry}
The asymmetry index ($A$) needs to be calculated first, because it sets the exact centre of the galaxy to be used in the following steps. For each possible centre (i.e. for each pixel in a box of side $r_{\rm P}/8$ around the \textsc{SExtractor} centre), we obtain an \textquotedblleft aperture image\textquotedblright{ }by masking the pixels outside the $1.5 r_{\rm P}$ circle/ellipse. Each aperture image is used to estimate the value of $A$ through equation \ref{eq:A} (in this phase we neglect the background term, since it is almost independent of the the centre position): the new centre is set to the pixel that minimizes $A$. Each time that this minimization procedure shifts the centre, we recompute both $r_{\rm P}$ and the position angle, and repeat the above procedure using the new parameters.
After a stable centre is found, the value of $A$ from the above procedure needs to be corrected for the background term in equation \ref{eq:A}. Details about this correction are given in Section \ref{app:pycasso-corrections}.

In contrast with \citet{conselice2003} and \citet{vikram2010}, we do not attempt to estimate the centre position with sub-pixel accuracy. This is because translations by a fraction of a pixel (and rotations by angles that are not multiples of $90^\circ$) require image interpolations that tend to smooth (and degrade) the original image. Our decision is supported by the comparison of the errors introduced by this smoothing with those due to the limited precision in the centre determination described in Appendix \ref{app:image-degradation}.

\subsubsection[]{Clumpiness}\label{app:pycasso-clumpiness}
\begin{figure}
\includegraphics[width=0.46\textwidth]{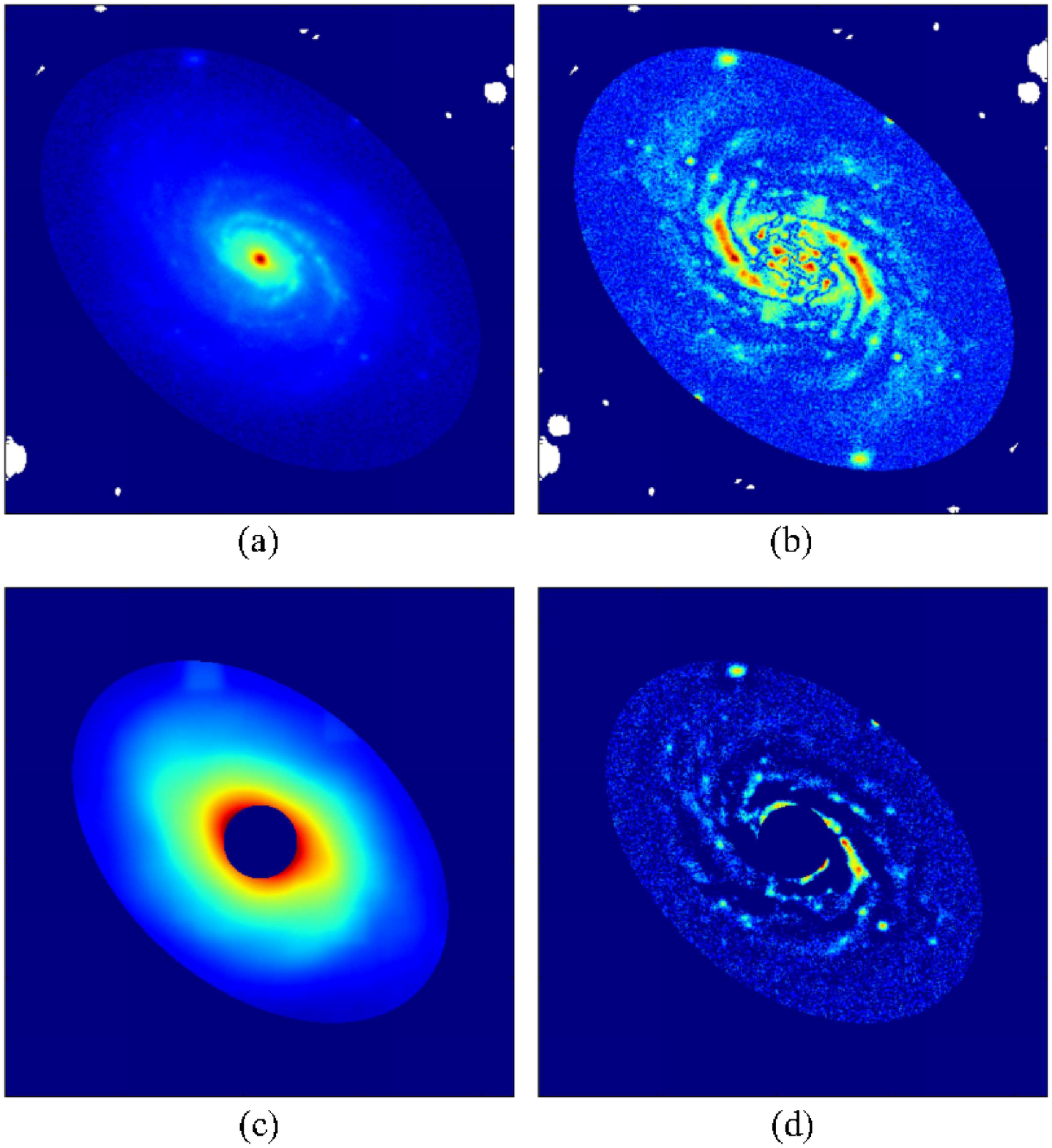}
\caption[]{Examples of images produced by \textsc{PyCASSo} while estimating the $CAS$ indexes: (a) aperture image enclosing the elliptical area of the galaxy; (b) residual after the subtraction of the $180^{\circ}$ rotated image (used for the computation of the asymmetry); (c) smoothing of the aperture image with a top hat filter (for the clumpiness computation); (d) residual image after the subtraction of the smoothed image (for the clumpiness computation). The dark-blue area around the ellipse of the galaxy is masked because it is outside of the $1.5r_{\rm P}$ limit. In images (c) and (d) there is also a central mask excluding the bulge/nucleus contribution. White areas show the masks applied by the software for removing foreground stars or other contaminating sources. Color normalizations are different between the four images.}
\label{fig:pycasso-cas}
\end{figure}

For the clumpiness ($S$) computation, we create a copy of the aperture image, and we smooth it with a top-hat filter of width $0.25 r_{\rm P}$ \citep{conselice2003}, so that the blurring scale is a fixed fraction of the galaxy size. The  smoothed image $I_S$ is subtracted from the original aperture image $I$ and, according to equation \ref{eq:S}, the intensities of all the {\it positive} pixels\footnote{The central $0.25 r_{\rm P}$ circular part of the galaxy is excluded from this computation \citep{conselice2003} because it might be contaminated (e.g. by an AGN). Furthermore, the \citet{conselice2003} procedure establishes that all the pixels where the subtraction of the smoothed image gives a negative result should be forced to 0.} of the residual image are summed, and this sum is multiplied by 10 and normalized by the cumulative intensity of the same pixels in the aperture image (see Figure \ref{fig:pycasso-cas}). In symbols (see also equation \ref{eq:S}),
\begin{equation}
	\begin{split}
	&S = 10 \, \frac{\sum_{i,j} [ I(i,j) - I_{S}(i,j) ] }{ \sum_{i,j} I(i,j) } \\
	 \forall (&i,j) \,\, {\rm such} \,\,{\rm that} \,\, I(i,j) > I_{S}(i,j) \, ,
	\end{split}
\end{equation}

As in the case of the asymmetry $A$, the clumpiness $S$ needs to be corrected for the background contribution, whose estimation is described in the next subsection.

\subsubsection[]{Background corrections}\label{app:pycasso-corrections}
The simple subtraction of the pixel mode applied in the image preparation phase guarantees that the image-averaged background is close to 0, but does not take into account brightness gradients and the granularity of the image; since they can affect the values of $A$ and $S$, a further correction must be applied (see \citealp{conselice2003}). Therefore, \textsc{PyCASSo} computes the spurious asymmetry and clumpiness in a square box of sky near the galaxy, and corrects the previous estimates of these indexes. We adopt the following criteria to spot the best sky area: 
(i) the box must be representative of the background, so it must be as free as possible from sources (stars, galaxies etc.). We search only for boxes where 80 per cent of the pixels, at least, belong to background (that is, the pixel intensity $B(x,y)$ is between $-3\sigma$ and $\sigma$, where $\sigma$ is the background standard deviation). Boxes that do not satisfy this requirement are discarded.
(ii) The box must be as close as possible to the galaxy, because it has to map the local properties of the sky. Therefore, we start from the edge of the aperture image, and search for all the boxes that satisfy condition (i) and do not overlap with the galaxy. If \textsc{PyCASSo} does not find at least five valid boxes along the loop, it restarts the search on a wider ring.
(iii) The box size should be comparable with the galaxy size. We initially search for boxes with the same area as the aperture image. If \textsc{PyCASSo} finds less than five boxes that satisfy criteria (i) and (ii), it reduces the box size by 20 per cent and repeats the search from the beginning.
When a search for background boxes satisfies all these requirements, \textsc{PyCASSo} computes the asymmetry of each box, and chooses the box with the lowest $A$. On the same box it computes the background clumpiness. Both these values are normalized by the total intensity of the galaxy and linearly rescaled with respect to the galaxy area.

In contrast with other authors \citep{vikram2010}, we give more importance to the size of the box than to its proximity to the galaxy. This choice is motivated by a test (see Table \ref{table:background-corrections}), showing that the asymmetry correction depends on the size of the box, even after rescaling is taken into account. Therefore, it is important to select a box of size as similar as possible to the galaxy, keeping the rescaling factor close to unity. Small-area asymmetries, rescaled to a much larger size, generally underestimate the asymmetry correction. On the contrary, the clumpiness correction does not show any trend. 

\begin{table}
\centering
\begin{tabular}{c c c}
\textbf{Box side (px)} &	\textbf{$A_{\rm background}$} &	\textbf{$S_{\rm background}$} \\
\hline
30	&		0.097		&		 0.0012 \\
60	&		0.163		&		 0.0152 \\
100	&		0.182		&		 0.0008 \\
150	&		0.186		&		 0.0186 \\
200	&		0.195		&		 0.0066 \\
250	&		0.212		&		 0.0012 \\
300	&		0.218		&		 0.0005 \\
350	&		0.223		&		 0.0022 \\
\hline
\end{tabular}
\caption{Example of the relation between the box size and the background corrections. The first column shows the size in pixels of the box side, the second and the third give the asymmetry and clumpiness corrections, respectively. For this table we used the NGC 4123 data from the \citet{frei} catalogue, but we obtained similar results with all the $\sim 10$ tested sources. Both the asymmetry and the clumpiness corrections have been already rescaled to the galaxy size, which corresponds to a square box of side $\sim 380$ pixels.}
\label{table:background-corrections}
\end{table}

\subsubsection[]{Concentration}\label{app:pycasso-concentration}
The concentration index is the ratio between the $r_{80}$ and $r_{20}$ radii, that contain respectively the 80 per cent and the 20 per cent of the total flux of the galaxy (operatively defined as the count sum of all the valid pixels inside the aperture image). These radii are computed starting from the centre of the galaxy and considering larger and larger apertures, until the interior flux reaches respectively the 20 per cent and the 80 per cent of the total; then we compute $C$ as in equation \ref{eq:C}. In the innermost part of the galaxy, where the brightness profile varies steeply (especially for galaxies with small angular scale), it is important to compute the radii with an high precision level (fractions of pixel). \textsc{PyCASSo} achieves this result by oversampling the image, i.e. by choosing a refinement factor $ref$ (between 1 and 10), converting each pixel in the aperture image into a square of side $(ref \times ref)$ pixels, each one with intensity equal to $ref^{-2}$ times the original value (to preserve the total flux), and then computing $r_{20}$ and $r_{80}$ on this enlarged image.

Our concentration values on the \citet{frei} galaxy catalogue are consistent within $1 \sigma$ with those provided by other authors (\citealp{conselice2003}, \citealp{vikram2010}); however, we note that our values of $C$ tend to be lower (by about 6--9 per cent) than other estimates. 

\begin{table}
\centering
\begin{tabular}{c c c c}
 & \textbf{Per cent} & & \\
 &	\textbf{of the mode} &	\textbf{$r_{\rm P}$} & \textbf{\textsl{C}} \\
\hline
\textbf{NGC 4030} & 100 & 58  & 3.44 \\
 									& 99  & 61  & 3.52 \\
 								  & 98  & 65  & 3.61 \\
 								  & 97  & 69  & 3.69 \\
\textbf{NGC 3198} & 100 & 94  & 2.82 \\
 									& 99  & 112 & 3.01 \\
\hline
\end{tabular}
\caption{Examples of the variation of the petrosian radius $r_{\rm P}$ and the concentration index due the uncertaity in the background intensity. A few percent variation in the background estimate may induce variations in the $C$ index of up to 10 per cent. Both galaxies are taken from the \citet{frei} catalogue.
\citealp{conselice2003} estimated $C= 3.67$ (NGC 4030) and $C=3.01$ (NGC 3198).}
\label{background-test}
\end{table}

The concentration index is computed directly from the light profile of the galaxy, so it depends mainly on the background subtraction. From our tests on the \citet{frei} galaxies, we see that an inaccuracy of only one per cent in the background value can lead to an error of up to 10 per cent in the concentration index (see the example in Table \ref{background-test}).

Other authors use different methods for the background subtraction, like exploiting the \textsc{SExtractor} background map \citep{vikram2010}, or a fit with a polynomial function (\citealp{conselice2003}, \citealp{hernandez2005}). However, these techniques might produce a local overestimate\footnote{\textsc{SExtractor} splits the image and evaluates the background in each sub-image through of a sigma clipping. This technique gives good results for a non-uniform background. However, if a galaxy is quite extended, it leads to a local overestimate of the background (because some of the boxes are almost entirely filled by the object). Therefore, the \textsc{SExtractor} background describes very well the empty regions of the frame and a possible brightness gradient, but tends to follow the light distribution of the sources. The same effect occurs with polynomial fits, because they tend to follow the intensity peaks produced by the sources in the image.} of the background in the area covered by the galaxy, and in particular at its centre, thus increasing the value of $r_{\rm P}$. As a test, we analyzed images after subtracting the \textsc{SExtractor} background image (rather than the one calculated by \textsc{PyCASSo}), and found that $C$ increases by $\sim5$ per cent (on average), making our results fully comparable with those by \citet{vikram2010} and \citet{conselice2003}.

Because of these considerations about background subtraction, we decided to keep our procedure, which is less subject to subtle artifacts. We remind that the concentration index is not used for any merging criteria: these small inconsistencies with other authors do not alter the science results of this paper.

\subsubsection[]{Segmentation image, Gini coefficient and second order momentum of light} \label{app:pycasso-gini-momentum}
\begin{figure}
\includegraphics[width=0.46\textwidth]{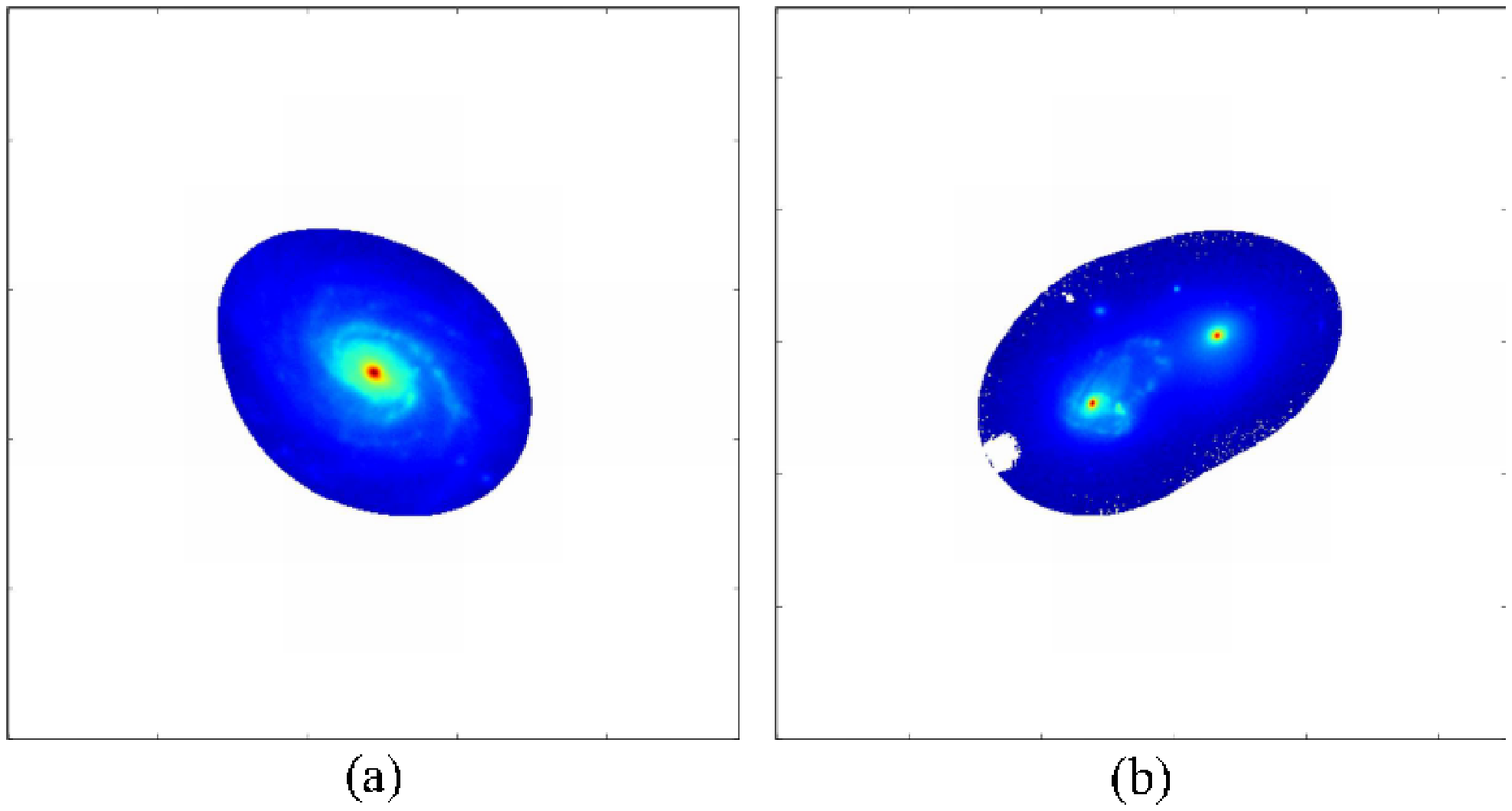}
\caption[]{Example of segmentation maps: the white region was removed because it does not satisfy the brightness lower limit, or contains sources not connected with the central galaxy. The map can assume an irregular shape that fits the galaxy outline. In the case of galaxy pairs (b), if the brightness distribution does not drop under $I_{\rm p}$ between one source and the other, they both are included in the segmentation map.}
\label{fig:pycasso-seg-map}
\end{figure}

The Gini coefficient and the second order momentum of light are not related to any of the $CAS$ indexes. They rely on another definition of the centre and of the area of the galaxy, defined through the segmentation map \citep{lotz2004}. First of all, by using elliptical apertures, we compute the mean intensity $I_{\rm P}$ at the Petrosian semi-major axis $a_{\rm P}$, and we convolve the image with a Gaussian filter of width $a_{\rm P}/5$. This step increases the $S/N$ ratio of the outer region of the galaxy, facilitating the identification of low surface brightness features. The segmentaton map is extracted from the original image, using only those pixels that, in the blurred image, satisfy the relation $I_{\rm P}\leq I \leq I_{\rm adj}+10\sigma_{\rm adj}$ (where $I$ is the pixel intensity, while $I_{\rm adj}$ and $\sigma_{\rm adj}$ are, respectively, the median and the standard deviation of the adjacent pixels), and that are topologically connected with the main body of the galaxy. The continuity requirement is quite weak, allowing the segmentation map to assume a very irregular shape and to follow the galaxy outline (see Figure \ref{fig:pycasso-seg-map}). 

From the segmentation map we derive the Gini coefficient by sorting pixels by decreasing intensity, and calculating $G$ as in equation \ref{eq:G}.

The computation of the momentum $M_{20}$ is more complicated: it requires to select a new centre of the galaxy by minimizing the value of the total second order momentum $M_{\rm tot}$. We start this search from the $CAS$ centre, and follow the same kind of procedure described in Section \ref{app:pycasso-asymmetry}.
When we find a stable centre, we sort the pixels by decreasing intensity and we compute $M_{20}$ using equation \ref{eq:M20}.
\\

\textsc{PyCASSo} saves the $CASGM$ indexes and all the other computed quantities (radii, position angles, background properties, etc.) in two ascii files. It saves also a series of control images (e.g. the figures in this Appendix) intended to help the user to verify the correctness of the analysis.

If the galaxy is too small, the resolution might be insufficient for the $CASGM$ analysis. Resolution and $S/N$ limits have been studied by \citep{lotz2004}: they found that $G$, $M_{20}$ and $C$ are reliable within $10$ per cent for images with an average $S/N$ ratio (per pixel) $\langle S/N \rangle \geq 2$; while $A$ and $S$ decrease sistematically with increasing $S/N$ (but variations are $\Delta A <0.1$ and $\Delta S < 0.2$ even for $\langle S/N \rangle \geq 5$). Low resolution has a stronger effect, because it flattens the brightness profile of the galaxy, increasing both the Petrosian radius and the angular size of the segmentation map. Usually, $G$, $A$ and $S$ are stable for spatial resolutions of 1 kpc or better, while $C$ and $M_{20}$ show a deviation greater than 15 per cent for spatial resolutions worse than 0.5 kpc, because the nuclei are no more resolved.
Beside these limits, we introduce a size requirement: the galaxy must have a Petrosian semi-major axis $a_{\rm P}>10$ pixels, otherwise the analysis may not be reliable, especially for what concerns the clumpiness, the Gini coefficient and the second order moment of light, because they rely on a further smoothing of the image.

\section[]{Image degradation}\label{app:image-degradation}
Quantitative analysis can be distorted even by small image degradations. Therefore, it is important to use procedures that minimize image alterations. The most common image-altering operations performed within the $CASGM$ analysis are translations and rotations.

Translations involving a shift by an integer number of pixels do not degrade image quality, because they simply move intensities from one pixel coordinate to another pixel coordinate. Instead, shifts by a fractional number of pixels require an interpolation, i.e. a weighted average on the values of neighbouring pixels. This has a smoothing effect whose relevance depends on the number of pixels involved (2 for a translation along one axis; 4, for a translation along both axis) and to the pixel weights (e.g., a shift by 0.1 pixel introduces less smoothing than a shift by 0.5 pixel).

Most of the rotations suffer from the same problem, since the original and final grids are not superimposable, and an interpolation/weighted average is applied. Rotations by angles multiple of $90^{\circ}$ exactly centred on a pixel are exceptions, since they can be obtained without altering the pixel values (by using image reflections and transpositions): this is particularly important for the $180^\circ$ rotation required for the computation of $A$, that can be achieved without introducing any image degradation.

Usually, the $CASGM$ index most affected by unwanted smoothings is $S$, but in some cases also $r_{\rm P}$ may be overestimated (because the averaging makes the light profile of the galaxy flatter), influencing also all the other parameters of the $CAS$ system.

\citet{conselice2003} and \citet{vikram2010} are not very restrictive in this respect: for example, they commonly use fractions of pixels. While this approach probably suffers from image degradation, it also presents some advantages: for example, it allows more precise (to the level of $\sim 0.1$ pixel rather than $\sim 0.5$ pixel) determinations of galaxy centres, that can lead to better values of $A$.

For this reason, we carried out tests comparing the loss of precision due to uncertainties in the centre positions, and the one due to the smoothing associated with translations and rotations. 
The former was evaluated by measuring the asymmetry also in the eight pixels around the centre, and computing the difference $\delta A_1$ between the asymmetry calculated in the real centre (we remind that the centre was chosen by minimizing $A$) and the lowest asymmetry of the neighbouring pixels.
The latter was evaluated by looking at the difference $\Delta A_2$ in the asymmetry of the same galaxy before and after two consecutive translations (in opposite directions, so that the image should return to its original position) by $0.5$ pixels along each axis.
We compared the two errors on the galaxies of the \citet{frei} catalogue: on average, $\Delta A_1/A \simeq 0.056$, while $\Delta A_2/A \simeq 0.127$, i.e. more then double\footnote{Had we performed a single ``forward'' translation, $\Delta A_2$ would be reduce by a factor $\sim 1.5$.}.

In the light of the results of the above test, we decided that \textsc{PyCASSo} should minimize image-degradation effects by applying only integer translations, and computing $180^\circ$ rotations through reflections.

\label{lastpage}


\begin{thebibliography}{99}
\bibitem[\protect\citeauthoryear{Abraham et al.}{1996}]{abraham1996} Abraham R.G., Tanvir N.R., Santiago B.X., Ellis R.S., Glazebrook K., van den Bergh S., 1996, MNRAS, 279, 47L
\bibitem[\protect\citeauthoryear{Abraham et al.}{2003}]{abraham2003} Abraham R., van den Bergh S., Nair P., 2003, ApJ, 588, 218
\bibitem[\protect\citeauthoryear{Alonso et al.}{2007}]{alonso07} Alonso M. S., Lambas D. G., Tissera P., Coldwell G., 2007, MNRAS, 375, 1017
\bibitem[\protect\citeauthoryear{Barnes \& Hernquist}{1991}]{barnes1991} Barnes J.E., Hernquist L.E., 1991, ApJ, 370, 65
\bibitem[\protect\citeauthoryear{Barnes}{1988}]{barnes88} Barnes J. E., 1988, ApJ, 331, 699
\bibitem[\protect\citeauthoryear{Barton, Geller \&{} Kenyon}{Barton et al.}{2000}]{barton00} Barton E. J., Geller M. J., Kenyon S. J., 2000, ApJ, 530, 660
\bibitem[\protect\citeauthoryear{Bell \&{} de Jong}{2001}]{belldejong2001} Bell E.F., de Jong R.S., 2001, ApJ, 550, 212
\bibitem[\protect\citeauthoryear{Bershady et al.}{2000}]{bershady2000} Bershady M., Jangren A., Conselice C., 2000, AJ, 119, 2645
\bibitem[\protect\citeauthoryear{Bertin \& Arnouts}{1996}]{bertin1996} Bertin E., Arnouts S., 1996, A\&AS, 317, 393
\bibitem[\protect\citeauthoryear{Best et al.}{2006}]{best06} Best P., Kaiser C. R., Heckman T. M., Kauffmann G., 2006, MNRAS, 368, L67
\bibitem[\protect\citeauthoryear{Bournaud et al.}{2005}]{bournaud2005} Bournaud F., Combes F., Jog C.J., Puerari I., 2005, A\&A, 438, 570
\bibitem[\protect\citeauthoryear{Bournaud et al.}{2011}]{bournaud11}Bournaud et al. 2011, ApJ, 730, 4
\bibitem[\protect\citeauthoryear{Blanton et al.}{2003}]{blanton2003} Blanton M. et al., 2003, ApJ, 594, 186
\bibitem[\protect\citeauthoryear{Blanton \& Roweis}{2007}]{blantonroweis2007} Blanton M.R., Roweis S., 2007, AJ, 133, 734
\bibitem[\protect\citeauthoryear{Byun \& Freeman}{1995}]{byun1995} Byun Y.I., Freeman K.C., 1995, ApJ, 448, 563
\bibitem[\protect\citeauthoryear{Churazov et al.}{2001}]{churazov01} Churazov E., Bruggen M., Kaiser C. R., Bohringer H., Forman W., 2001, ApJ, 554, 261
\bibitem[\protect\citeauthoryear{Cisternas et al.}{2011}]{cisternas2011} Cisternas et al., 2011, ApJ, 726, 57
\bibitem[\protect\citeauthoryear{Coldwell \&{} Lambas}{2006}]{coldwelllambas06} Coldwell G. V., Lambas D. G., 2006, MNRAS, 371, 786
\bibitem[\protect\citeauthoryear{Conselice, Bershady \&{} Jangren}{Conselice et al.}{2000}]{conselice00} Conselice Ch. J., Bershady M. A., Jangren A., 2000, ApJ, 529, 886
\bibitem[\protect\citeauthoryear{Conselice}{2000b}]{conselice2000b} Conselice C.J., Bershady M.A., Gallagher J.S., 2000b, A\&A, 354, 21L
\bibitem[\protect\citeauthoryear{Conselice}{2003}]{conselice2003} Conselice C., 2003, ApJS, 147, 1
\bibitem[\protect\citeauthoryear{Conselice}{2006}]{conselice2006} Conselice C., 2006, ApJ, 638, 686
\bibitem[\protect\citeauthoryear{Croton et al.}{2006}]{croton06} Croton, D.J., Springel, V., White, S.D.M., et al. 2006, MNRAS, 365, 11 
\bibitem[\protect\citeauthoryear{Cusumano et al.}{2010}]{palermo} Cusumano G. et al., 2010, A\&A, 524, A64 
\bibitem[\protect\citeauthoryear{Dahari}{1984}]{dahari84} Dahari O., 1984, AJ, 89, 966
\bibitem[\protect\citeauthoryear{Dahari}{1985}]{dahari85} Dahari O., 1985, ApJS, 57, 643
\bibitem[\protect\citeauthoryear{Darg et al.}{2010}]{darg10} Darg D. W. et al., 2010, MNRAS, 401, 1552
\bibitem[\protect\citeauthoryear{De Propris et al.}{2007}]{depropris2007} De Propris R., Conselice C.J., Liske J., Driver S.P., Patton D.R., Graham A.W., Allen P.D., 2007, ApJ, 666, 212
\bibitem[\protect\citeauthoryear{Dekel et al.}{2009}]{dekel09}Dekel A. et al. 2009, Nat, 457, 451
\bibitem[\protect\citeauthoryear{Di Matteo et al.}{2005}]{dimatteo2005} Di Matteo T., Springel V., Hernquist L., 2005, Nat, 433, 604
\bibitem[\protect\citeauthoryear{Dunlop et al.}{2003}]{dunlop03} Dunlop J. S., McLure R. J., Kukula M. J., Baum S. A., O'Dea C. P., Hughes D. H., 2003, MNRAS, 340, 1095
\bibitem[\protect\citeauthoryear{Ellison et al.}{2008}]{ellison08} Ellison S. L., Patton D. R., Simard L., McConnachie A. W., 2008, AJ, 135, 1877
\bibitem[\protect\citeauthoryear{Ellison et al.}{2011}]{ellison11} Ellison S. L., Patton D. R., Mendel J. T., Scudder J. M., 2011, MNRAS, 418, 2043
\bibitem[\protect\citeauthoryear{Ferrarese \&{} Ford}{2005}]{ferrareseford05} Ferrarese L., Ford H., 2005, Space Science Reviews, 116, 523
\bibitem[\protect\citeauthoryear{Ferrarese \&{} Merritt}{2000}]{ferraresemerritt00} Ferrarese L., Merritt D., 2000, ApJ, 539, L9
\bibitem[\protect\citeauthoryear{Frei et al.}{1996}]{frei} Frei Z.,  Guhathakurta P., Gunn J., Tyson J.A., 1996, AJ, 111, 174
\bibitem[\protect\citeauthoryear{Fuentes-Williams \&{} Stocke}{1988}]{fuentesstocke88} Fuentes-Williams Th., Stocke J. T., 1988, AJ, 96, 1235
\bibitem[\protect\citeauthoryear{Gabor et al.}{2009}]{gabor2009} Gabor J.M. et al., 2009, ApJ, 691, 705
\bibitem[\protect\citeauthoryear{Gebhardt et al.}{2000}]{gebhardt00}  Gebhardt K., et al., 2000, ApJ, 539, L13
\bibitem[\protect\citeauthoryear{Gehrels}{1986}]{gehrels1986} Gehrels N., 1986, ApJ, 303, 336
\bibitem[\protect\citeauthoryear{Gehrels et al.}{2004}]{gehrels2004} Gehrels N. et al., 2004, ApJ, 611, 1005
\bibitem[\protect\citeauthoryear{Gerhard}{1981}]{gerhard81} Gerhard O. E., 1981, MNRAS, 197, 179 
\bibitem[\protect\citeauthoryear{Glasser}{1962}]{glasser1962} Glasser G.J., 1962, J. Amer. Stat. Assoc. 57, 648, 654
\bibitem[\protect\citeauthoryear{Graham}{2012a}]{graham12a} Graham Alister W., 2012a, ApJ, 746, 113
\bibitem[\protect\citeauthoryear{Graham}{2012b}]{graham12b} Graham Alister W., 2012b, MNRAS, 422, 1586
\bibitem[\protect\citeauthoryear{Grogin et al.}{2005}]{grogin05} Grogin N. A. et al., 2005, ApJ, 627, L97
\bibitem[\protect\citeauthoryear{H\"aring \&{} Rix}{2004}]{haringrix04} H\"aring N., Rix H.W., 2004, ApJ, 604, L89
\bibitem[\protect\citeauthoryear{Heckman \&{} Kauffmann}{2011}]{heckmankauffmann11} Heckman T. M., Kauffmann G., 2011, Sci, 333, 182
\bibitem[\protect\citeauthoryear{Hernandez-Toledo et al.}{2005}]{hernandez2005} Hernandez-Toledo H.M., Avila-Reese V., Conselice C.J., Puerari I., 2005, AJ, 129, 682
\bibitem[\protect\citeauthoryear{Hopkins et al.}{2006}]{hopkins2006} Hopkins P.F., Hernquist L.; Cox, T.J., Di Matteo T., Robertson B., Springel V., 2006, ApJS, 163, 1
\bibitem[\protect\citeauthoryear{Kauffmann \&{} Haehnelt}{2000}]{kauffmannhaehnelt00} Kauffmann G., Haehnelt M., 2000, MNRAS
\bibitem[\protect\citeauthoryear{Keel et al.}{1985}]{keel85} Keel W. C., Kennicutt R. C. Jr., Hummel E., van der Hulst J. M., 1985, AJ, 90, 708
\bibitem[\protect\citeauthoryear{Kendall \& Stuart}{1977}]{kendall1977} Kendall M., Stuart A., 1977, The advanced theory of statistics. Vol.1: Distribution theory, Wiley
\bibitem[\protect\citeauthoryear{Keres et al.}{2005}]{keres05} Keres D., Katz N., Weinberg D. H., Dav\'e R., 2005, MNRAS, 363, 2
\bibitem[\protect\citeauthoryear{Kocevski et al.}{2012}]{kocevski12} Kocevski D. D. et al., 2012, ApJ, 744, 148
\bibitem[\protect\citeauthoryear{Jogee et al.}{2009}]{jogee2009} Jogee S. et al., 2009, ApJ, 697, 1971
\bibitem[\protect\citeauthoryear{Koss et al.}{2010}]{koss2010} Koss M., Mushotzky R., Veilleux S., Winter L., 2010, ApJ, 716, 125
\bibitem[\protect\citeauthoryear{Koss et al.}{2011}]{koss11} Koss M. et al., 2011, ApJ, 739, 57
\bibitem[\protect\citeauthoryear{Koss et al.}{2012}]{koss2012} Koss M., Mushotzky R., Treister E., Veilleux S., Vasudevan R., Trippe M., 2012, ApJ, 746, 22
\bibitem[\protect\citeauthoryear{Koulouridis et al.}{2006}]{koulouridis06} Koulouridis E., Plionis M., Chavushyan V., Dultzin-Hacyan D., Krongold Y., Goudis C., 2006, ApJ, 639, 37 
\bibitem[\protect\citeauthoryear{Lake \&{} Dressler}{1986}]{lakedressler86} Lake G., Dressler A., 1986, ApJ, 310, 605 
\bibitem[\protect\citeauthoryear{Le F\'evre}{2000}]{lefevre2000} Le F\'evre et al., 2000, MNRAS, 311, 565
\bibitem[\protect\citeauthoryear{Li et al.}{2008}]{li08} Li C., Kauffmann G., Heckman T. M., White S. D. M., Jing Y. P., 2008, MNRAS, 385, 1915
\bibitem[\protect\citeauthoryear{Liu, Shen \&{} Strauss}{2012}]{liu12} Liu X., Shen Y., Strauss M. A., 2012, ApJ, 745, 94
\bibitem[\protect\citeauthoryear{Lotz et al.}{2004}]{lotz2004} Lotz J.M., Primack J., Madau P., 2004, AJ, 613, 262
\bibitem[\protect\citeauthoryear{Lotz et al.}{2008}]{lotz2008b} Lotz J.M., Jonsson P., Cox T.J., Primack I.R., 2008, MNRAS, 391, 1137
\bibitem[\protect\citeauthoryear{Magorrian et al.}{1998}]{magorrian98} Magorrian J., et al., 1998, AJ, 115, 2285
\bibitem[\protect\citeauthoryear{Mapelli, Moore \&{} Bland-Hawthorn}{Mapelli et al.}{2008}]{mapelli08} Mapelli M., Moore B., Bland-Hawthorn J., 2008, MNRAS, 388, 697	
\bibitem[\protect\citeauthoryear{Marconi \&{} Hunt}{2003}]{marconihunt03} Marconi A., Hunt, L. K., 2003, ApJ, 589, L21
\bibitem[\protect\citeauthoryear{Marconi et al.}{2004}]{marconi04} Marconi A. et al., 2004, MNRAS, 351, 169
\bibitem[\protect\citeauthoryear{McNamara \&{} Nulsen}{2007}]{mcnamaranulsen07} McNamara B. R., Nulsen P. E. J., 2007, Annu. Rev. Astron. Astrophys., 45, 117
\bibitem[\protect\citeauthoryear{Menci et al.}{2008}]{menci2008} Menci N., Fiore F., Puccetti S., Cavaliere A., 2008, ApJ, 686, 219
\bibitem[\protect\citeauthoryear{Merloni \& Heinz}{2012}]{merloni12} Merloni, A., \& Heinz, S.\ 2012, arXiv:1204.4265 
\bibitem[\protect\citeauthoryear{Miller et al.}{2003}]{miller03} Miller Ch. J., Nichol R. C., G\'omez P. L., Hopkins A. M., Bernardi M., 2003, ApJ, 597, 142 
\bibitem[\protect\citeauthoryear{Miller \&{} Smith}{1980}]{millersmith80} Miller R. H., Smith B. F., 1980, ApJ, 235, 421
\bibitem[\protect\citeauthoryear{Mirabel}{2001}]{mirabel01} Mirabel I. F., 2001, ApSSS, 277, 371
\bibitem[\protect\citeauthoryear{Negroponte \&{} White}{1983}]{negrowhite83} Negroponte J., White S. D. M., 1983, MNRAS, 205, 1009
\bibitem[\protect\citeauthoryear{Paturel et al.}{2003}]{paturel2003} Paturel G., Petit C., Prugniel Ph., Theureau G., Rousseau J., Brouty M., Dubois P., Cambrésy L., 2003, A\&A, 412, 45
\bibitem[\protect\citeauthoryear{Patton et al.}{2000}]{patton2000} Patton D.R., Carlberg R.G., Marzke R.O., Pritchet C.J., da Costa L.N., Pellegrini P.S., 2000, ApJ, 536, 153
\bibitem[\protect\citeauthoryear{Patton et al.}{2002}]{patton2002} Patton D.R. et al., 2002, ApJ, 565, 208
\bibitem[\protect\citeauthoryear{Patton \& Atfield}{2008}]{patton2008} Patton D.R., Atfield J.E., 2008, ApJ, 685, 235
\bibitem[\protect\citeauthoryear{Peng et al.}{2002}]{peng2002} Peng C. Y., Ho L.C., Impey C.D., Rix H.R., 2002, AJ, 124, 266
\bibitem[\protect\citeauthoryear{Petrosian}{1976}]{petrosian1976} Petrosian V., 1976, ApJ, 209, L1
\bibitem[\protect\citeauthoryear{Petrosian}{1982}]{petrosian82} Petrosian A. R., 1982, Astrofizika, 18, 548
\bibitem[\protect\citeauthoryear{Pierce et al.}{2007}]{pierce2007} Pierce C.M. et al., 2007, ApJ, 660, 19
\bibitem[\protect\citeauthoryear{Rafanelli, Violato \&{} Baruffolo}{Rafanelli et al.}{1995}]{rafanelli95}Rafanelli P., Violato M., Baruffolo A., 1995, AJ, 109, 1546
\bibitem[\protect\citeauthoryear{Ramos Almeida et al.}{2011}]{ramosalmeida11} Ramos Almeida C., Tadhunter C. N., Inskip K. J., Morganti R., Holt J., Dicken D., 2011, MNRAS, 410, 1550
\bibitem[\protect\citeauthoryear{Reichard et al.}{2008}]{reichard08} Reichard T. A., Heckman T. M., Rudnick G., Brinchmann J., Kauffmann G., 2008, ApJ, 677, 186
\bibitem[\protect\citeauthoryear{Sales et al.}{2012}]{sales12}Sales L.V. et al. 2012, MNRAS, 3041S
\bibitem[\protect\citeauthoryear{Sancisi et al.}{2008}]{sancisi08} Sancisi R., Fraternali F., Oosterloo T., van der Hulst Th. 2008, The Astronomy and Astrophysics Review, 15, 189
\bibitem[\protect\citeauthoryear{Sanders}{1988}]{sanders1988} Sanders D.B., Soifer B.T., Elias J.H., Madore B.F., Matthews K., Neugebauer G., Scoville, N.Z., 1988, ApJ, 325, 74
\bibitem[\protect\citeauthoryear{Scarlata}{2007}]{scarlata2007} Scarlata C. et al., 2007, ApJS, 172, 494
\bibitem[\protect\citeauthoryear{Schawinski et al.}{2006}]{schawinski06} Schawinski K. et al., 2006, Nat, 442, 888
\bibitem[\protect\citeauthoryear{Schawinski et al.}{2007}]{schawinski07} Schawinski K. et al., 2007, MNRAS, 382, 1415
\bibitem[\protect\citeauthoryear{Schawinski et al.}{2009}]{schawinski09} Schawinski K. et al., 2009, ApJ, 690, 1672
\bibitem[\protect\citeauthoryear{Schawinski et al.}{2010}]{schawinski10} Schawinski K., Dowlin N., Thomas D., Urry C. M., Edmondson E., 2010, ApJ, 714, L108
\bibitem[\protect\citeauthoryear{Schmitt}{2001}]{schmitt01} Schmitt H. R., 2001, AJ, 122, 2243
\bibitem[\protect\citeauthoryear{Serber et al.}{2006}]{serber06} Serber W., Bahcall N., M\'enard B., Richards G.,  2006, ApJ, 643, 68
\bibitem[\protect\citeauthoryear{Silverman et al.}{2011}]{silverman11} Silverman J. D., 2011, ApJ, 743, 2
\bibitem[\protect\citeauthoryear{Smirnova, Moiseev \&{} Afanasiev}{Smirnova et al.}{2010}]{smirnova10}Smirnova A. A., Moiseev A. V., Afanasiev V. L., 2010, MNRAS, 408, 400
\bibitem[\protect\citeauthoryear{Springel, Di Matteo \&{} Hernquist}{Springel et al.}{2005}]{springel05} Springel V., Di Matteo T., Hernquist L., 2005, MNRAS, 361, 776
\bibitem[\protect\citeauthoryear{Vikram et al.}{2010}]{vikram2010} Vikram V., Wadadekar Y., Kembhavi A.K., Vijayagovindan, G.V., 2010, MNRAS, 409, 1379
\bibitem[\protect\citeauthoryear{Virani, De Robertis \&{} VanDalfsen}{Virani et al.}{2000}]{virani00} Virani S. N., De Robertis M. M., VanDalfsen M. L., 2000, AJ, 120, 1739
\bibitem[\protect\citeauthoryear{Volonteri \& Bellovary}{2012}]{volonteri12} Volonteri, M., \& Bellovary, J.\ 2012, Reports on Progress in Physics, 75, 124901 
\bibitem[\protect\citeauthoryear{Waskett et al.}{2005}]{waskett05} Waskett T. J., Eales S. A., Gear W. K., McCracken H. J., Lilly S., Brodwin M., 2005, MNRAS, 363, 801
\bibitem[\protect\citeauthoryear{White}{1978}]{white78} White S. D. M. 1978, MNRAS, 184, 185
\end{thebibliography}
\end{document}